\documentclass[PAPER, atlasdraft=false, UKenglish, cernpreprint, texmf, orcidlogo]{atlasdoc}
\usepackage[subcaption]{atlaspackage}
\usepackage{atlasbiblatex}

\usepackage{atlasphysics}
\usepackage{multirow}

\graphicspath{{logos/}{figures/}}

\usepackage{ANA-STDM-2018-43-PAPER-defs}

\AtlasTitle{Measurements of the production cross-section for a $Z$ boson in association with $b$- or $c$-jets in proton--proton collisions at $\sqrt{s} = 13$ TeV with the ATLAS detector}

\AtlasAbstract{%
This paper presents a measurement of the production cross-section of a $Z$ boson in association with $b$- or $c$-jets, in proton--proton collisions at $\sqrt{s} = 13$ TeV with the ATLAS experiment at the Large Hadron Collider using data corresponding to an integrated luminosity of 140 fb$^{-1}$.
Inclusive and differential cross-sections are measured for events containing a $Z$ boson decaying into electrons or muons and produced in association with at least one $b$-jet, at least one $c$-jet, or at least two $b$-jets
with transverse momentum $p_\textrm{T} > 20$ GeV and rapidity $|y| < 2.5$.
Predictions from several Monte Carlo generators based on next-to-leading-order matrix elements interfaced
with a parton-shower simulation, with different choices of flavour schemes for initial-state partons, are compared with the
measured cross-sections. The results are also compared with novel predictions, based on infrared and collinear safe jet flavour dressing algorithms. Selected $Z + \ge 1~c$-jet observables, optimized for sensitivity to intrinsic-charm, are compared with benchmark models with different intrinsic-charm fractions.
}

\AtlasRefCode{STDM-2018-43}

\PreprintIdNumber{CERN-EP-2024-081}

\arXivId{2403.15093[hep-ex]}

\AtlasJournalRef{JHEP 07 (2020) 044}
\AtlasDOI{10.1007/JHEP07(2020)044}

\hypersetup{pdftitle={ATLAS document},pdfauthor={The ATLAS Collaboration}}

\begin{document}

\maketitle

\tableofcontents
\clearpage

\section{Introduction}

The measurement of the production rate of a \Zbos in  association with jets generated from heavy-flavour quarks, namely \cquarks\ and \bquarks,\footnote{
Throughout the paper it is implicitly assumed that \cquark\ and \bquark\ refer to both quark and antiquark.}
leading to \Zcjets and \Zbjets final states, in proton--proton ($pp$) collisions provides an important test of perturbative quantum chromodynamics (pQCD) and of the proton internal structure.
Measurements of \Zcjets\ and \Zbjets\ production provide a benchmark
to probe the modelling from Monte Carlo (MC) generators used to estimate the background contribution of these processes to other topologies, such as Higgs boson measurements or searches for new physics which often feature
resonant decay into a pair of \bquarks.

State-of-the-art MC predictions for \Zbjets\ and \Zcjet\ production are available for different flavour and mass schemes and typically based on next-to-leading-order (NLO) matrix elements combined with a parton shower (PS).
In the 4-flavour number scheme (4FS), \bquarks can only be generated perturbatively from an explicit gluon splitting ($g\to b\bar b$) in the partonic matrix element, are typically treated as massive, and do not contribute to the parton distribution functions (PDFs) of the proton.
This method does not resum logarithms of the type $\alpha_{s}\ln(Q^{2}/m_b^{2})$, where $Q$ is the hard process scale and $m_b$ the quark mass, making it suitable for processes close to the \bquark mass scale.
In the 5-flavour number scheme  (5FS), the \bquark density is allowed in the initial state via a \bquark PDF, allowing the resummation of the $\alpha_{s}\ln(Q^{2}/m_b^{2})$ terms to all orders in $\alpha_{s}$. This method is suitable for predictions where the scale of the process is much larger than the \bquark mass, by treating the \bquark as massless.
The ambiguity among the schemes is an intrinsic property of the calculation and is expected to reduce with the inclusion of higher-order perturbative corrections~\cite{Vbteo3}.
While infrared-safe algorithms are routinely used for inclusive jets at the Large Hadron Collider (LHC), a variety of such algorithms has now also been developed for flavoured jets~\cite{Gauld:2022lem,Gauld:2023zlv,Czakon_2023,Caletti_2022}.

Inclusive and differential cross-sections of \Zbjets production were measured in proton--anti-proton collisions at the centre-of-mass energy of $\sqrt{s} = 1.96$\,\TeV by the CDF and D0 experiments~\cite{Aaltonen:2008mt, Abazov:2013uza}. Measurements of \Zbjets processes were performed at the
LHC in \pp collisions at $\sqrt{s} = 7$\,\TeV by the ATLAS, CMS and LHCb experiments~\cite{STDM-2012-15, Chatrchyan:2014dha, Chatrchyan:2013zja, Chatrchyan:2012vr, LHCb:2014ydc}, at $\sqrt{s} = 8$\,\TeV by the CMS experiment~\cite{Khachatryan:2016iob} and at $\sqrt{s} = 13$\,\TeV by the ATLAS experiment with a partial dataset in both resolved~\cite{STDM-2017-38} and boosted~\cite{STDM-2017-37} regimes, and by the CMS experiment with the full Run~2 dataset~\cite{CMS-SMP-20-015}.

Predictions for \Zcjets production are  derived either in a 3-flavour scheme (3FS) or in a 4/5FS\@.
Similarly to the discussion for \Zbjets production, the two approaches differ by the presence of
resummation of $\alpha_{s}\ln(Q^{2}/m_c^{2})$ terms. The 4/5FS calculations that resum these logarithms
are suitable for calculations where the scale of the process is much larger than the mass of \cquark, treating them as massless.
In this context, the hypothesis of a valence-like, or intrinsic, component of charm quarks in a proton was first proposed nearly 40 years ago~\cite{Brodsky:1980pb,Brodsky:1981se}.
Although major PDF fitters also provide PDF sets including intrinsic charm (IC)~\cite{Ball:2016neh,Hou:2017khm}, its existence and amount is still subject of a long-standing debate, lacking experimental observations.
It was shown that the measurement of photon and \Zbos production in association with \cjets\ in the forward rapidity region is sensitive to the effect of IC~\cite{Bednyakov:2013zta,Beauchemin:2014rya}.
Inclusive and differential cross-sections of \Zcjets\ production were measured in $\sqrt{s} = 13$\,\TeV \pp collisions by CMS with a partial dataset~\cite{CMS-SMP-19-011}. The LHCb experiment has measured \Zcjets\ cross-sections in the forward region~\cite{LHCb:2021stx}
demonstrating a sizeable excess over predictions made with no-IC PDF in the region of high \Zbos rapidity, which was later interpreted as evidence of the IC in the proton~\cite{Ball:2022qks}.

This paper presents measurements of the inclusive and differential production cross-sections of a \Zbos, decaying into electrons or muons, in association with at least one \cjet, at least one \bjet or at least two \bjets using  the full \pp collision dataset collected by the ATLAS experiment at $\sqrt{s} = 13$\,\TeV and corresponding to \Lumi.
Measurements of the inclusive \Zonec, \Zoneb and \Ztwob cross-sections in a fiducial phase space are performed.
Differential measurements are performed for observables selected for their sensitivity to pQCD, PDF models and MC generator validation, as detailed in Table~\ref{tab:int:diffxsec}.
In particular:
the  transverse momentum, \pT, of the \Zbos and of the highest-\pT (leading) heavy-flavour jet are able to probe both the pQCD predictions and MC modelling for \Zoneb or \Zonec production;
\DeltaRZb is chosen as it is sensitive to the presence of additional radiation in the event, a substantially different feature of 4FS versus 5FS calculations;
the $\xF(c)$ and \RptZ observables, for  \Zonec events, are sensitive to PDF models and focus on IC sensitivity;
finally, \mbb  and  \DeltaPhibb  are sensitive to various features of MC generators, in particular $g\to b\bar{b}$ modelling, in the \Ztwob final state, relevant for Higgs boson measurements or searches for new physics.

\begin{table}[h!]
\begin{center}
\caption{List of observables used to perform differential cross-section measurements.}\label{tab:int:diffxsec}
\begin{tabular}{l l l}
\toprule
Final state  & Observable 	& Notation      \\
\midrule
\Zoneb      & $p_{\textrm{T}}$  of the leading $b$-jet 	& $p_{\mathrm{T,}b}^0$    \\
& $p_{\textrm{T}}$   of the $Z$ boson    	& $p_{\textrm{T}}(Z)$         \\
& $\Delta \tilde R = \sqrt{(\Delta\phi)^2+ (\Delta y)^2}$ between the $Z$ boson and leading $b$-jet,  	& $\Delta \tilde R_{Zb}$     \\
& where $\Delta\phi$ ($\Delta y$) is the azimuthal angle (rapidity) difference & \\
\midrule
\Zonec      & $p_{\textrm{T}}$  of the leading $c$-jet	& $p_{\mathrm{T,}c}^0$     \\
& $p_{\textrm{T}}$   of the $Z$ boson  	& $p_{\textrm{T}}(Z)$         \\
& Feynman-$x$ variable $x_F=2|p_z(c)|/\sqrt{s}$~\cite{Feynman:1969ej}        & $\xF(c)$             \\
& Cross-section ratio of $p_{\textrm{T}}(Z)$ in $|y(Z)|<1.2$ and $|y(Z)|>1.2$ 		& R($p_{\textrm{T}}(Z)$)           \\
\midrule
\Ztwob      & Invariant mass of the two leading $b$-jets        	& $m_{bb}$            \\
& Azimuthal angle difference between the two leading $b$-jets  	& $\Delta\phi_{bb}$  \\
\bottomrule
\end{tabular}
\end{center}
\end{table}

In comparison with the previous ATLAS measurement~\cite{STDM-2017-38}, more extreme regions of the phase space are explored. The higher integrated luminosity contributes to increase the number of selected events for rare final states with \cjets or two \bjets. Advances in jet reconstruction and $b$-tagging and the new data-driven methodologies developed for the estimate of main backgrounds allow a reduction of the leading experimental and modelling uncertainties, and thus increase the sensitivity to the signal processes.
In addition, the measurement of \RptZ, i.e. the  \ptZ differential cross-section ratio of forward over central  \Zonec\ events,\footnote{The separation of the central and forward  \Zonec\ event categories for $|y(Z)|=1.2$ is optimized on the basis of  sensitivity to IC models and statistical precision of the expected measurement.} allows further cancellation of systematic uncertainties enhancing sensitivity to IC effects.
Finally, recent advances in the theory sector permit a comparison of data with MC predictions using next-to-leading-order matrix elements, expected to provide a better description of the processes.

The strategy of the measurement is the following.
Events with a \Zbos candidate decaying into an electron or muon pair produced in association with at least one or at least two jets identified as containing a $b$- or a $c$-hadron (flavour-tagged) are selected based on variables measured at detector level.
Top quark pair and multijet backgrounds are  estimated via data-driven techniques,
while the remaining non-\Zjets backgrounds (mainly dibosons) are estimated via MC simulations.
Selected events containing processes with a genuine \Zbos are categorized as \Zbjets, \Zcjets\ and \Zljets\
through a likelihood fit on a flavour-sensitive observable (``flavour fit'').
The fit is done separately for events with $\ge 1$ flavour-tagged jet and with $\ge 2$ flavour-tagged jets.
Fits are performed in each bin of each observable, using the data to constrain the shape of that observable and the normalisation.
Separate electron and muon channel distributions are fitted simultaneously.
The non-signal \Zjets components, normalised in the flavour fit, are then subtracted from the data together with the other electroweak (EWK) and top backgrounds.
After background subtraction, data are unfolded to particle level using an iterative approach, in a fiducial phase-space close to the detector-level selection.
Unfolding inputs are obtained by summing electron and muon-channel  distributions.
The detector-level systematic uncertainties are propagated by performing the flavour fits and unfolding separately for each systematic variation in a correlated way.
Unfolded distributions of $Z\,+ \ge 1$ or $\ge 2$~\bjets and \Zonec are compared with available signal predictions from
state-of-the-art generators and from NNLO fixed-order predictions. \Zonec measurements are finally compared with several PDF sets with different contributions from IC.

This paper is organized as follows. The ATLAS detector is described in Section~\ref{sec:detector}, and details of the data sample and the MC simulations are provided in Section~\ref{sec:datasets}.
The object definitions and the event selection at detector level are presented in Section~\ref{sec:selection}. Backgrounds that do not contain a real \Zbos are estimated via MC simulations or via data-driven techniques, while backgrounds containing a real \Zbos
and jets are estimated with a fit to data distributions sensitive to the flavour of the jet; both are described in Section~\ref{sec:bkgEstimation}. Distributions of the kinematic variables are presented in Section~\ref{sec:distribution}. After background subtraction, the data are unfolded to particle level in a fiducial phase space, as detailed in Section~\ref{sec:unfolding}. Systematic uncertainties in the unfolded data are discussed in Section~\ref{sec:uncertainties}. The results are presented in Section~\ref{sec:results}, and conclusions are drawn in Section~\ref{sec:conclusion}.


%

\newcommand{\AtlasCoordFootnote}{%
ATLAS uses a right-handed coordinate system with its origin at the nominal interaction point (IP)
in the centre of the detector and the \(z\)-axis along the beam pipe.
The \(x\)-axis points from the IP to the centre of the LHC ring,
and the \(y\)-axis points upwards.
Polar coordinates \((r,\phi)\) are used in the transverse plane,
\(\phi\) being the azimuthal angle around the \(z\)-axis.
The pseudorapidity is defined in terms of the polar angle \(\theta\) as \(\eta = -\ln \tan(\theta/2)\).
Angular distance is measured in units of \(\Delta R \equiv \sqrt{(\Delta\eta)^{2} + (\Delta\phi)^{2}}\).}

\section{ATLAS detector}
\label{sec:detector}

The ATLAS detector~\cite{PERF-2007-01} at the LHC covers nearly the entire solid angle around the collision point.\footnote{\AtlasCoordFootnote}
It consists of an inner tracking detector surrounded by a thin superconducting solenoid, electromagnetic and hadronic calorimeters,
and a muon spectrometer incorporating three large superconducting air-core toroidal magnets.

The inner-detector system (ID) is immersed in a \qty{2}{\tesla} axial magnetic field
and provides charged-particle tracking in the range \(|\eta| < 2.5\).
The high-granularity silicon pixel detector covers the vertex region and typically provides four measurements per track,
the first hit generally being in the insertable B-layer (IBL) installed before Run~2~\cite{ATLAS-TDR-19,PIX-2018-001}.
It is followed by the SemiConductor Tracker (SCT), which usually provides eight measurements per track.
These silicon detectors are complemented by the transition radiation tracker (TRT),
which enables radially extended track reconstruction up to \(|\eta| = 2.0\).
The TRT also provides electron identification information
based on the fraction of hits (typically 30 in total) above a higher energy-deposit threshold corresponding to transition radiation.

The calorimeter system covers the pseudorapidity range \(|\eta| < 4.9\).
Within the region \(|\eta|< 3.2\), electromagnetic calorimetry is provided by barrel and
endcap high-granularity lead/liquid-argon (LAr) calorimeters,
with an additional thin LAr presampler covering \(|\eta| < 1.8\)
to correct for energy loss in material upstream of the calorimeters.
Hadronic calorimetry is provided by the steel/scintillator-tile calorimeter,
segmented into three barrel structures within \(|\eta| < 1.7\), and two copper/LAr hadronic endcap calorimeters.
The solid angle coverage is completed with forward copper/LAr and tungsten/LAr calorimeter modules
optimised for electromagnetic and hadronic energy measurements respectively.

The muon spectrometer (MS) comprises separate trigger and
high-precision tracking chambers measuring the deflection of muons in a magnetic field generated by the superconducting air-core toroidal magnets.
The field integral of the toroids ranges between \num{2.0} and \qty{6.0}{\tesla\metre}
across most of the detector.
Three layers of precision chambers, each consisting of layers of monitored drift tubes, cover the region \(|\eta| < 2.7\),
complemented by cathode-strip chambers in the forward region, where the background is highest.
The muon trigger system covers the range \(|\eta| < 2.4\) with resistive-plate chambers in the barrel, and thin-gap chambers in the endcap regions.

The luminosity is measured mainly by the LUCID--2~\cite{LUCID2} detector that records Cherenkov light produced in the quartz windows of photomultipliers located close to the beampipe.

Events are selected by the first-level trigger system implemented in custom hardware,
followed by selections made by algorithms implemented in software in the high-level trigger~\cite{TRIG-2016-01}.
The first-level trigger accepts events from the \qty{40}{\MHz} bunch crossings at a rate below \qty{100}{\kHz},
which the high-level trigger further reduces in order to record complete events to disk at about \qty{1}{\kHz}.

A software suite~\cite{ATL-SOFT-PUB-2021-001} is used in data simulation, in the reconstruction
and analysis of real and simulated data, in detector operations, and in the trigger and data acquisition
systems of the experiment.

%
%

%
%
%
%
%
%
%
%
%

%
%
%

%

%
%
%
%
%
%
%
%
%
%
%
%
%
%
%
%
%
%
%


%

\section{Data and simulated event samples}
\label{sec:datasets}

\subsection{Data sample description}

The data used in this measurement were collected with the ATLAS detector at the LHC between 2015 and 2018 in \pp collisions at $\sqrt s = 13$\,\TeV (LHC Run~2).
Crossings of proton bunches occurred every 25~ns, the collisions achieved a peak instantaneous luminosity of \mbox{$2.1\times 10^{34}$~cm$^{-2}$s$^{-1}$}, and the mean number of \pp interactions per bunch crossing (pile-up) was $\langle \mu \rangle = 34$.

The recorded data correspond to a total integrated luminosity of $140$~\ifb with an uncertainty of 0.83\%~\cite{DAPR-2021-01}, obtained using the LUCID-2 detector~\cite{LUCID2} for the primary luminosity measurements, complemented by measurements using the inner detector and calorimeters. Data quality requirements are applied to ensure that all detector components were in good operating conditions.

The candidate events are selected by either a single-electron or single-muon trigger with minimum transverse momentum threshold, quality and isolation requirements~\cite{TRIG-2018-05,TRIG-2018-01,TRIG-2019-03}.
The \pT threshold in 2015 is 24 (20)\,\GeV for the electrons (muons), satisfying loose isolation requirements. Due to the higher instantaneous luminosity in 2016--2018, the \pT threshold is increased to 26\,\GeV for both the electrons and the muons,
and more restrictive isolation requirements are imposed on both leptons along with more restrictive identification requirements for electrons.
Triggers with higher \pT thresholds but with no isolation requirement or with loosened identification criteria were also used to increase the efficiency.

\subsection{Simulated event samples for signal and background processes}\label{subsec:MC}

MC simulations are used to describe signal events, to estimate the contribution of background processes, to unfold the data yield to the particle level, to estimate systematic uncertainties and in comparison with the unfolded data distributions. Generated events are processed using a full detector simulation~\cite{SOFT-2010-01} based on $\GEANT$~\cite{Agostinelli:2002hh}, for the detector response to final-state particles, and then reconstructed using the same algorithms as the data.
To account for pile-up, multiple overlaid inelastic \pp collisions are simulated with \PYTHIA[8.186]~\cite{Sjostrand:2007gs} using the A3 tune~\cite{ATL-PHYS-PUB-2016-017} and the \NNPDFtwoLO\ PDF set~\cite{Ball:2012cx}. The distribution of the average number of interactions per bunch crossing in the simulations is weighted to reflect the one in data.

An overview of all signal and background processes and the generators used for the production of nominal results is given in Table~\ref{tab:generators}.

\begin{table}[!tb]
\begin{center}
\caption{Summary of signal and background MC samples. The generator programs used in the simulation are listed in the second column.
The QCD order in the calculation of the matrix elements (ME) is indicated in the third column, where $n$p denotes the number of real parton emissions, and FS refers to the flavour scheme used.
The precision in QCD calculation for the inclusive production cross-section ($\sigma_\mathrm{prod}$) is provided in the forth column.}
\footnotesize
\renewcommand{\arraystretch}{1.1}
\begin{tabular}{c | c c c}
\toprule
Process   & Generator	&  Order of pQCD  in ME  (FS) 		& Order $\sigma_\mathrm{prod}$ calculation \\%
\midrule
$Z \to \ell \ell$ & \textsc{MGaMC+Py8 FxFx}    &  0--3p NLO (5FS) 			&  \\
$Z \to \ell \ell$ & \textsc{Sherpa 2.2.11}    &  0--2p NLO, 3--5p LO (5FS) 		&  \\
\midrule
$\ttbar$ 	& NNLO+NNLL	\\
single top ($s$/$t$/$Wt$-channel) 	& \textsc{Powheg+Py8}  	&  NLO	& NLO  \\
\midrule
$qg/q\bar{q} \to VV \to \ell\ell/\ell\nu/\nu\nu +q\bar{q}$   & \sherpa~2.2.1 		& 1p NLO, 2--3p LO 	& NLO \\
\midrule
$qq \to ZH \to \ell\ell/\nu\nu + b\bar{b}$  & \textsc{Powheg+Py8}     & NLO 	&NNLO (QCD), NLO (EWK) \\
$gg \to ZH \to \ell\ell/\nu\nu + b\bar{b}$  & \textsc{Powheg+Py8}     & NLO 	& NLO+NLL \\
\bottomrule
\end{tabular}
\label{tab:generators}
\end{center}
\end{table}

Inclusive \Zbos production in association with both light- and heavy-flavour jets is simulated using the \MGNLO{} v2.6.5~\cite{Alwall:2014hca} generator, combined with \pythia v8.245~\cite{Sjostrand:2014zea} parton shower using the \textsc{FxFx}~\cite{Frederix:2012ps} merging procedure (\mgfxfx). This program generates matrix elements for \Zbos production, with up to three additional partons in the final state at NLO accuracy in the strong coupling constant~\cite{PMGR-2021-01}. The showering and subsequent hadronisation is performed using \pythia v8.245 with the A14 tune~\cite{ATL-PHYS-PUB-2014-021}, using the \NNPDFtwoLO PDF set with $\alpha_{s}=0.130$. The different jet multiplicities are merged using the \fxfx{} NLO matrix element and parton shower merging prescription. The \mgfxfx\ calculation uses the 5FS with massless $b$- and $c$-quarks at the matrix element level, and massive quarks in the \pythia~8 parton shower. At the generation level, the jet momentum is required to be at least 10~\GeV. The PDF set used for event generation is \NNPDF[3.1luxQED] with $\alpha_{s}=0.118$~\cite{NNPDF:2017mvq}. The merging scale is set to $Q_{\textrm{cut}}=20$~\GeV. \mgfxfx\ \Zjets\ samples are used for the nominal signal and background determination, unfolding of the data distributions, to estimate the uncertainties and are compared with the final cross-section measurements.
$Z(\to \tau\tau)$+jets background samples are also modelled by \mgfxfx.

Additional $Z(\to \ell \ell)+$jets ($\ell = e, \mu$) samples are simulated with the ATLAS configuration of \SHERPA ~\cite{PMGR-2021-01}, which computes NLO-accurate matrix elements for up to two partons and LO-accurate matrix elements for up to five partons with the Comix~\cite{Gleisberg:2008fv} and \textsc{OpenLoops}~\cite{Buccioni:2019sur,Cascioli:2011va,Denner:2016kdg} libraries. The default \sherpa parton shower~\cite{Schumann:2007mg} based on a Catani--Seymour dipole factorization and the cluster hadronisation model~\cite{Winter:2003tt}
is used. \SHERPA performs a 5FS calculation with massless $b$- and $c$-quarks at the ME level, and massive quarks in the parton shower. The Hessian \NNPDFthreeNNLO set of PDFs~\cite{Ball:2014uwa} is used, calculated at next-to-next-to-leading order (NNLO) in QCD with $\alpha_{s}=0.118$.
The samples are produced using a dedicated set of parameters developed by the \sherpa authors. The different jet multiplicities are merged using the \MEandPSatNLO prescription~\cite{Hoeche:2011fd,Hoeche:2012yf,Catani:2001cc,Hoeche:2009rj}, with merging scale $Q_{cut}$ set to 20~\GeV.
The samples provide an alternative in the modelling of $Z(\to \ell \ell)+$jets processes to \mgfxfx. In particular they are used to determine the uncertainty on the \Zjets background, to evaluate the modelling uncertainty in the unfolding procedure and finally in comparison with the measured data.

The $t\bar{t}$ simulations use the \powhegbox~v2~\cite{Frixione:2007nw,Nason:2004rx,Frixione:2007vw,Alioli:2010xd} generator at NLO with the \NNPDF[3.0nlo]~\cite{Ball:2014uwa} PDF set and the \hdamp\ parameter\footnote{
The \hdamp\ parameter is a resummation damping factor and one of the parameters that control the matching of \powheg matrix elements to the parton shower and thus effectively regulates the high-\pt\ radiation against which the \ttbar\ system recoils.} set to 1.5~\mtop~\cite{ATL-PHYS-PUB-2016-020}.
The events are interfaced to \PYTHIA[8.230]~\cite{Sjostrand:2014zea}, using the A14 tune~\cite{ATL-PHYS-PUB-2014-021} and the \NNPDF[2.3lo]
set of PDFs~\cite{Ball:2012cx}. This sample is normalized to the NNLO cross-section, including the resummation of next-to-next-to-leading logarithmic (NNLL) soft gluon terms, calculated with \textsc{Top}$++$2.0~\cite{Beneke:2011mq,Cacciari:2011hy,Baernreuther:2012ws,Czakon:2012zr,Czakon:2012pz,Czakon:2013goa,Czakon:2011xx}.

Single top quark production, in association with $W$ bosons ($tW$) as well as in the $s$- and $t$- channels, is also simulated using \powhegbox~v2 at NLO
using the \NNPDF[3.0nlo] PDF set interfaced with \PYTHIA[8.230] using the \NNPDFtwoLO PDF set.
To remove overlap with the $t\bar{t}$ sample, the diagram-removal~\cite{Frixione:2008yi} procedure was used.

Semileptonic diboson ($VV$) final states, where one $W$ or \Zbos decays leptonically and the other decays hadronically, were simulated with \sherpa v.2.2.1~\cite{Bothmann:2019yzt},
with NLO MEs for up to one parton emission and LO MEs for up to three parton emissions calculated using the \NNPDFthreeNNLO PDF set. The ME calculations are matched to the \sherpa parton shower using the MEPS@NLO prescription.

Simulated events for $qq\to ZH(\to \ell \ell +b\bar{b})$ and $gg\to ZH(\to \ell \ell+b\bar{b})$ processes in inclusive production or in association with one jet at NLO are generated with the \powhegbox~v2 interfaced with \PYTHIA[8.230] generator with the \NNPDFthreeNNLO PDF set. The samples include all final states where the Higgs boson decays into $b\bar{b}$ and the \Zbos in a leptonic final state.
The mass of the Higgs boson is set to 125~\GeV and the $H\to b\bar{b}$ branching ratio is set to 58\%. The $qq\to ZH(\to \ell \ell +b\bar{b})$ cross-section is calculated at NNLO (QCD) and NLO (EWK), while $gg\to ZH(\to \ell \ell+b\bar{b})$ cross-section is calculated at NLO+NLL (QCD).
Contributions from $t\bar{t}Z$ and $tZ$ processes are found to be negligible and are not included.

\FloatBarrier

\subsection{Theoretical predictions}
\label{sec:datasets:theory}
In addition to particle-level predictions from the fully simulated \mgfxfx and \SHERPA samples described above, the measured unfolded cross-sections are compared with several theoretical predictions, in order to test the sensitivity to different flavour schemes in the matrix element calculation, to the intrinsic-charm component in the proton PDF and to higher orders in QCD calculation.

The generator predictions, all based on \mgamc, and their respective settings are summarized in Table~\ref{tab:PdfSummary}.

To test the 4FS predictions, a sample of $Z+b\bar b$ events generated by \MGNLO{} v3.5.3 is produced using the 4FS with massive \bquarks\ at NLO with two partons in the matrix element.
It uses the 4-flavour \NNPDFthreeoneNLO PDF set (with only perturbative charm contribution) with $\alpha_s = 0.118$; the sample is combined with \pythia v8.310 for the parton shower and hadronisation.
To provide a 5FS reference for these predictions, another sample is produced with \MGNLO{} v2.7.3 in the 5FS, where \Zbjets and \Zcjets production is modelled with
massless quarks in the matrix element, generated at NLO accuracy with one parton. It uses the \NNPDFtwoNLO PDF with $\alpha_s = 0.118$.
The showering and subsequent hadronisation is performed with \pythia v8.243 where \bquarks\ are treated as massive.

For the 3FS prediction of \Zcjets production, a $Z+c\bar c$ event sample is generated with \MGNLO{} v3.5.3 at NLO with two partons,
treating \cquarks\ as massive in the matrix element. It uses the 3-flavour \NNPDFthreeoneNLO PDF set (with only perturbative charm contribution) with $\alpha_s = 0.118$.
The showering and subsequent hadronisation is done with \pythia v8.310 with massive \cquarks.

Finally, a variety of predictions testing several IC models were produced using the PDF reweighting technique
according to the prescriptions in Ref.~\cite{Butterworth:2015oua}. The standard \mgfxfx sample is used as a baseline for PDF reweighting.
One possibility to include the IC is provided by the NNPDF group and allows a fraction of IC already in the nominal
\NNPDF[3.1luxQED] PDF used
in the \mgfxfx sample. A later analysis still allows it in the nominal \NNPDF[4.0] PDF sets~\cite{NNPDF:2021njg}.
Moreover, fits including the LHCb $Z+$charm measurement~\cite{LHCb:2021stx} and EMC data~\cite{EuropeanMuon:1982xfn} on top
of the baseline NNPDF dataset have been performed~\cite{Ball:2022qks}. In the latter case, the NNPDF group has claimed a $3\sigma$
evidence of presence of the intrinsic charm in the proton.
Predictions for the \Zcjets processes are produced using the nominal \NNPDFfourNNLO PDF and the one including the LHCb and EMC data.
To produce a no-IC reference to those, another prediction is made using the \NNPDF[4.0 NNLO (pch)] PDF that includes only
perturbative charm contribution.

Two predictions are produced using the CTEQ-TEA PDF sets including IC contributions under various assumptions~\cite{Guzzi:2022rca}.
One of them is made using a PDF based on the CT18NNLO fit implementing the BHPS3 model of IC~\cite{Hou:2017khm,Blumlein:2015qcn}.
The other one uses a CT18NNLO variant including IC following a meson-baryon model based on the effective-mass (MBME)
quark model~\cite{Hobbs:2013bia}. As a reference for these, a no-IC prediction is made using the nominal CT18NNLO PDF that
includes only perturbative charm contribution.

Lastly, two predictions are made using earlier PDFs from CT14 fits that include a fixed amount of IC\@. They are based on
the CT14NNLO baseline fit and correspond to BHPS1 and BHPS2 models that include IC with an
average fraction of proton momentum carried by intrinsic \cquarks\ of 0.6\% and 2.1\%,
respectively~\cite{Hou:2017khm}. A no-IC prediction with the nominal CT14NNLO PDF is also provided as no-IC reference.

\begin{table}[!tb]
\begin{center}
\caption{Summary of all MC prediction along with the flavour schemes (FS) and PDFs they use. PDF ID numbers are given according to LHAPDF~\cite{Butterworth:2014efa} numbering scheme.}\label{tab:PdfSummary}
\footnotesize
\begin{tabular}{l|c|l|r}
\toprule
Generator/settings & Flav. scheme & PDF & LHAPDF ID \\
\midrule
\multicolumn{4}{c}{Main MC samples}\\
\mgfxfx & 5FS & NNPDF3.1~(NNLO) LuxQED & 325100 \\
\SHERPA & 5FS & NNPDF3.0~(NNLO)        & 303200 \\
\midrule
\multicolumn{4}{c}{Predictions to test various flavour schemes}\\
\mgamc & 5FS & NNPDF2.3~(NLO) & 229800 \\
\mgamc $Zbb$  & 4FS & NNPDF3.1~(NLO) \textsc{pch} & 321500 \\
\mgamc $Zcc$ & 3FS & NNPDF3.1~(NLO) \textsc{pch} & 321300 \\
\midrule
\multicolumn{4}{c}{Intrinsic charm (IC) predictions}\\
\multirow{9}{*}{\textsc{MGaMC+Py8 FxFx}} & \multirow{9}{*}{5FS} & NNPDF4.0~(NNLO) \textsc{pch} (no IC) & 332100 \\
& & NNPDF4.0~(NNLO) & 331100 \\
& & NNPDF4.0~(NNLO) EMC+LHCbZc & -- \cite{Ball:2022qks}
\\
\cmidrule{3-4}
& & CT18~(NNLO) (no IC) & 14000 \\
& & CT18FC -- CT18 BHPS3  & 14087 \\
& & CT18FC -- CT18 MCM-E  & 14093 \\
\cmidrule{3-4}
& & CT14~(NNLO) (no IC) & 13000 \\
& & CT14~(NNLO)IC -- BHPS1 & 13082 \\
& & CT14~(NNLO)IC -- BHPS2 & 13083 \\
\midrule
\multicolumn{4}{c}{Fixed-order predictions~\cite{Gauld:2023zlv}}\\
NLO  & 5FS & PDF4LHC21 & 93000 \\
NNLO  & 5FS & PDF4LHC21 & 93000 \\
\bottomrule
\end{tabular}

\end{center}
\end{table}

In addition, fixed-order NLO and NNLO QCD \Zjets\ predictions with massless quarks are obtained for signal comparisons, following the numerical set-up of~\cite{Gauld:2023zlv}.
Notably, the flavour of the jets are assigned following an infrared and collinear (IRC) safe algorithm referred to as
the flavour-dressing algorithm~\cite{Gauld:2022lem}.\footnote{Numerically, the main difference of this definition is that jets containing a $b\bar b$ or a $c\bar c$ pair (i.e. double tagged jets) are considered flavour-less, a contribution which becomes more relevant in the high \ptZ regime.}


%

\section{Object definitions and event selections}
\label{sec:selection}

In this measurement, events are required to have a signature consistent with a \Zbos,
decaying into two electrons or two muons, in association with at least one or at least two \bjets or at least one \cjet. Events are selected if they are recorded during stable beam conditions and if they satisfy detector and data-quality requirements~\cite{DAPR-2018-01}.
Candidate events are required to have a primary vertex (PV), defined as the vertex with the highest sum of track $p_{T}^{2}$, with at least two associated tracks measured in the ID, each with $\pt > 500$~\MeV \cite{ATL-PHYS-PUB-2015-026}.

Electron candidates are reconstructed from ID tracks coming from the PV, which are matched to clusters of energy deposited in the electromagnetic calorimeters. Electrons must fulfil the following PV conditions: the longitudinal impact parameter ($z_0$) is required to satisfy  $|z_0 \sin(\theta)|<0.5$~mm, where $\theta$ is the angle of the track with respect to the beam-line, and the transverse impact parameter significance ($d_0/\sigma_{d_0}$) is required to satisfy $d_0/\sigma_{d_0}< 5$. Electrons must satisfy requirements on the shape of the electromagnetic shower in the calorimeter, track quality and track-cluster matching, using a likelihood-based identification with a `Tight' working-point~\cite{EGAM-2018-01}. Electrons must be isolated, passing the `Tight-Var' working point, built from tracking and calorimeter information, with a energy-dependent variable cone~\cite{EGAM-2018-01}. Electron are required to have $\pt > 27$~\GeV\ and $|\eta| < 2.47$. Candidates in the transition region between the
barrel and end-caps of the electromagnetic calorimeters ($1.37 < |\eta|  < 1.52$) are excluded.

Muon candidates are identified by matching ID tracks coming from the PV to either full tracks or track segments reconstructed in the muon spectrometer.
The candidates must pass the following PV requirements: the longitudinal impact parameter must satisfy $|z_0\sin(\theta)| < 0.5$~mm and the transverse impact parameter significance must satisfy $|d_0|/\sigma(d_0) < 3$. Muons are required to fulfil `Medium' identification requirements~\cite{PERF-2015-10} based on quality criteria applied to the inner-detector and muon-spectrometer tracks. Muons must be isolated, passing the `FCTight' working point, built from tracking and calorimeter information, with a \pT-dependent variable cone~\cite{MUON-2018-03}. Muon candidates are required to have $\pt > 27$~\GeV\ and $|\eta| < 2.5$.

Hadronic jets are reconstructed using the anti-$k_t$ algorithm~\cite{Cacciari:2008gp} implemented in the \textsc{FastJet} package~\cite{Cacciari:2011ma}, with radius parameter $R = 0.4$, from particle-flow objects ~\cite{PERF-2015-09}.
Jets are calibrated using a simulation-based calibration scheme, followed by in situ corrections to account for differences between simulation and data~\cite{PERF-2016-04}. To eliminate jets coming from pile-up vertices, jets with $\pT< 60$~\GeV\ and $|\eta|<2.4$ are required to have a significant fraction of their tracks with origin compatible with the PV, as defined by a jet vertex tagger (JVT) discriminant~\cite{PERF-2014-03}.
Selected jets must have $\pt > 20$~\GeV\ and rapidity $|y|<2.5$.

Jets containing heavy-flavour hadrons are identified using a cut on the DL1r $b$-tagging discriminant~\cite{FTAG-2019-07}, a deep-learning multivariate algorithm trained using information about tracks and the secondary vertexes within a jet. Jets are selected if they pass the 85\% working point, which corresponds to an efficiency of 85\% in selecting jets containing a $b$-hadron, of 30\% for jets containing a $c$-hadron (and no $b$-hadrons) and of 2.5\% for jets containing only light hadrons.
Flavour-tagged jets are required to have $\pT > 20$~\GeV and pseudorapidity $|\eta|<2.5$.

Electrons, muons and jets are reconstructed and identified independently. An overlap-removal procedure is then applied to uniquely identify these objects in an event. Preselected jets with a high probability to be initiated by an electron or a radiated photon, such that $\Delta R$ between the jet and a lepton is smaller than 0.2, are removed. In a second step, leptons closer than $\Delta R = 0.4$ to any remaining jet are removed.

The missing transverse momentum $\vec{p}_\mathrm{T}^\mathrm{miss}$, with magnitude \MET,
is defined as the negative vector sum of the transverse momentum of all identified hard physics objects (electrons, muons, jets), as well as an additional track-based soft term accounting for the contribution of unclustered particles,  as in Ref.~\cite{ATLAS-CONF-2018-023}.

Events are required to have exactly two leptons of the same flavour ($ee$ or $\mu \mu$) and of opposite charge with dilepton invariant mass in the range of \SI{76}{\GeV}$<m_{\ell\ell}<$\SI{106}{\GeV}. At least one of the lepton candidates is required to match the lepton that triggered the event.
To reject the large contribution from \ttbar background, events with \mbox{$\pT^{\ell\ell} < 150$~\GeV} must have $\MET<$~\SI{60}{\GeV}.
For events passing these selection criteria two signal regions are defined: those with at least one \ftag jet form the ``\ZOneTag'' region; and those with at least two \ftag jets form the ``\ZTwoTag'' region.

A summary of the object definitions and the event selections used in the analysis is given in Table~\ref{tab:ZBJetSelection}.

\begin{table}[!tb]
\begin{center}
\caption{Summary of object definitions and event selections defining the signal regions of the analysis at detector level.}
\footnotesize
\renewcommand{\arraystretch}{1.1}
\begin{tabular}{m{0.16\textwidth}|m{0.27\textwidth}|m{0.27\textwidth}}
\toprule
\multicolumn{3}{c}{Object definition}\\
\toprule
& Electron channel 			& Muon channel \\
\midrule
& Single electron trigger		& Single muon trigger\\
& Tight    							& Medium \\
& Isolated 							& Isolated \\
Leptons 	& $d_0/\sigma_{d_0}< 5$, $|z_0 \sin(\theta)|<0.5$~mm			& $d_0/\sigma_{d_0}< 3$, $|z_0 \sin(\theta)|<0.5$~mm \\
& $p_{\textrm{T}}$ $>$ 27~\GeV 														& $p_{\textrm{T}}$ $>$ 27~\GeV \\
& $|\eta| < 1.37$ or $1.52<|\eta| < 2.47$ 							& $|\eta| < 2.5$ \\
\midrule
Jets 		&  \multicolumn{2}{c}{ $p_{\textrm{T}}$   $>$ 20~\GeV and $|y| < 2.5$}\\
&  \multicolumn{2}{c}{  $\Delta R(\text{jet}, \ell)     > 0.4$}\\
\midrule
\multirow{2}{*}{Flavour-tagged jets}			& \multicolumn{2}{c}{ $p_{\textrm{T}}$ $>$ 20~\GeV and $|y|$ < 2.5}  \\
& \multicolumn{2}{c}{DL1r$@85\%$ }\\
\bottomrule

\end{tabular}

\vspace{0.5em}
\renewcommand{\arraystretch}{1.1}

\begin{tabular}{ m{0.25\textwidth}|m{0.45\textwidth} }
\toprule
\multicolumn{2}{c}{Event selection}\\
\midrule
Leptons    				& Exactly 2, same-flavour, opposite-charge\\
$m_{\ell\ell}$ 		& 76~GeV $< m_{\ell\ell} <$ 106~GeV	\\
$E_{\textrm T}^{\mathrm{miss}}$           			& $E_{\textrm T}^{\mathrm{miss}}$ $<$ 60~GeV if $p_{\textrm{T}}^{\ell\ell}<$ 150~GeV	\\
Flavour-tagged jets           	& $\geq 1$ or $\geq 2$ jets, DL1r$@85\%$ \\
\midrule
\multicolumn{2}{c}{Signal regions}\\
\midrule
\ZOneTag & $\geq 1$ flavour-tagged jets  \\
\ZTwoTag & $\geq 2$ flavour-tagged jets  \\
\midrule
\multicolumn{2}{c}{Rapidity regions}\\
\midrule
Central rapidity & $Z$ boson rapidity $|y(Z)| < 1.2$ \\
Forward rapidity & $Z$ boson rapidity $|y(Z)| \ge 1.2$ \\
\midrule
\bottomrule
\end{tabular}
\label{tab:ZBJetSelection}
\end{center}
\end{table}

\FloatBarrier

In MC samples, events passing the selection in Table~\ref{tab:ZBJetSelection} are furthermore categorised according to the flavour of the underlying hadron generating the jet.
First, the selected jets are classified as either $b$-, $c$- or light jets using cone-based criteria as follows.
They are labelled as \bjets if they lie within $\Delta R = 0.3$ of at least one $b$~hadron with $\pT>5$\,\GeV. If a $b$~hadron matches two jets, only the closest jet in $\Delta R$ is labelled as \bjet. Jets not identified as \bjets are considered to be \cjets if they lie within $\Delta R = 0.3$ of any $c$~hadron with $\pT>5$\,\GeV. All other jets are classified as light jets.
Then, simulated events are sequentially classified as follows.
If they have heavy-flavour jets and the leading one is a \bjet, they are classified as \Zb or \Zbb when there is exactly one
or more than one \bjet, respectively. If the leading heavy-flavour jet is a \cjet, they are classified as \Zc. Finally,
they are classified as \Zl when only light jets are present.

In the \ZOneTag signal region, \Zoneb and \Zonec analyses are performed.
In the former, the sum of \Zb and \Zbb samples is used to define the signal,
while the \Zjets{} background corresponds to the sum of the \Zc and \Zl samples.
In the latter, \Zc is used to define the signal, while the sum of the \Zb, \Zbb{} and \Zl{} samples is treated
as \Zjets background.
In the \ZTwoTag signal region, \Ztwob analysis is performed, where the signal is given by the \Zbb sample, while the sum of the \Zb, \Zc, and \Zl samples form the \Zjets background.


%

\section{Background estimation}
\label{sec:bkgEstimation}

The dominant background in both the \ZOneTag and \ZTwoTag signal regions is composed by events with a \Zbos produced in association
with jets where jet flavour is different from the one targeted in the measurement, i.e. light jets
and either \cjets for \Zoneb and \Ztwob measurements or \bjets for \Zonec measurement.
This is determined using a fit to data, as detailed in Section~\ref{sec:FlavFit}.
Dileptonic \ttbar events represent the second largest background,
especially important in \ZTwoTag region. Its contribution is estimated with a data-driven method, as explained in Section~\ref{sec:DataDrivenTTbar}.
Smaller background contributions from dibosons, $ZH$, single top quark and $Z \to \tau \tau$ productions are estimated using simulation, as described in Section~\ref{subsec:MC}.
Background contributions from multijet events are estimated with a data-driven technique and found to be negligible as described in Section~\ref{sec:MultiJetBackground}.

\subsection{Data-driven estimation of \ttbar\ contribution}
\label{sec:DataDrivenTTbar}

A control region enriched with \ttbar events is constructed using the same event selection of Table~\ref{tab:ZBJetSelection}, but requiring
opposite-flavour $e^\pm\mu^\mp$ final states, instead of same-flavour $e^+e^-$ and $\mu^+\mu^-$ ones, and a wider
dilepton mass window \mbox{71~\GeV $<\mll<$ 111~\GeV}.
This control region contains only percent-level contributions from single-top production events and other processes.
Examples of \ttbar\ control region distributions are shown in Figure~\ref{fig:ttbarCR}.

The prediction for the \ttbar\ distributions\footnote{This data-driven estimate automatically includes minor processes where \ttbar dileptonic events are produced in association with vector bosons.}
in the signal regions is obtained by subtracting all non-\ttbar\ simulated backgrounds from the data in the control region
and multiplying by $ee/e\mu$ and $\mu\mu/e\mu$ transfer factors.
The transfer factors are derived as the ratio of the simulated \ttbar distribution in the signal region to that in the control region for each observable.

All detector systematic uncertainties described in \secRef{sec:uncertainties} are propagated to the \ttbar estimate by
applying corresponding variation to the simulated \ttbar sample used for deriving the transfer factors, and to
the non-\ttbar\ simulated backgrounds in the control region.
The propagation of theoretical uncertainties associated with the  simulated \ttbar sample was found to have negligible effect.
An additional uncertainty is introduced to evaluate possible bias from the
extrapolation of \ttbar contribution to the signal region. It is evaluated in a validation region defined to be
similar to the signal region but requiring $\MET\geq 60$\,\GeV, irrespective of the value of $\pT^{\ell\ell}$, and
requiring the dilepton mass to satisfy 71~\GeV~$<$~\mll~$<$~76~\GeV or 106~\GeV~$<$~\mll~$<$~111~\GeV. The uncertainty
is computed as difference between the \ttbar estimate by extrapolation from the control region and the result of subtracting
simulated non-\ttbar contributions in the validation region. This difference reaches 10\% for low transverse momenta of \Zbos and of leading
tagged jet and is mostly consistent with zero elsewhere.

\begin{figure}[!tb]
\begin{subfigure}{0.49\textwidth}
\includegraphics[width=\textwidth]{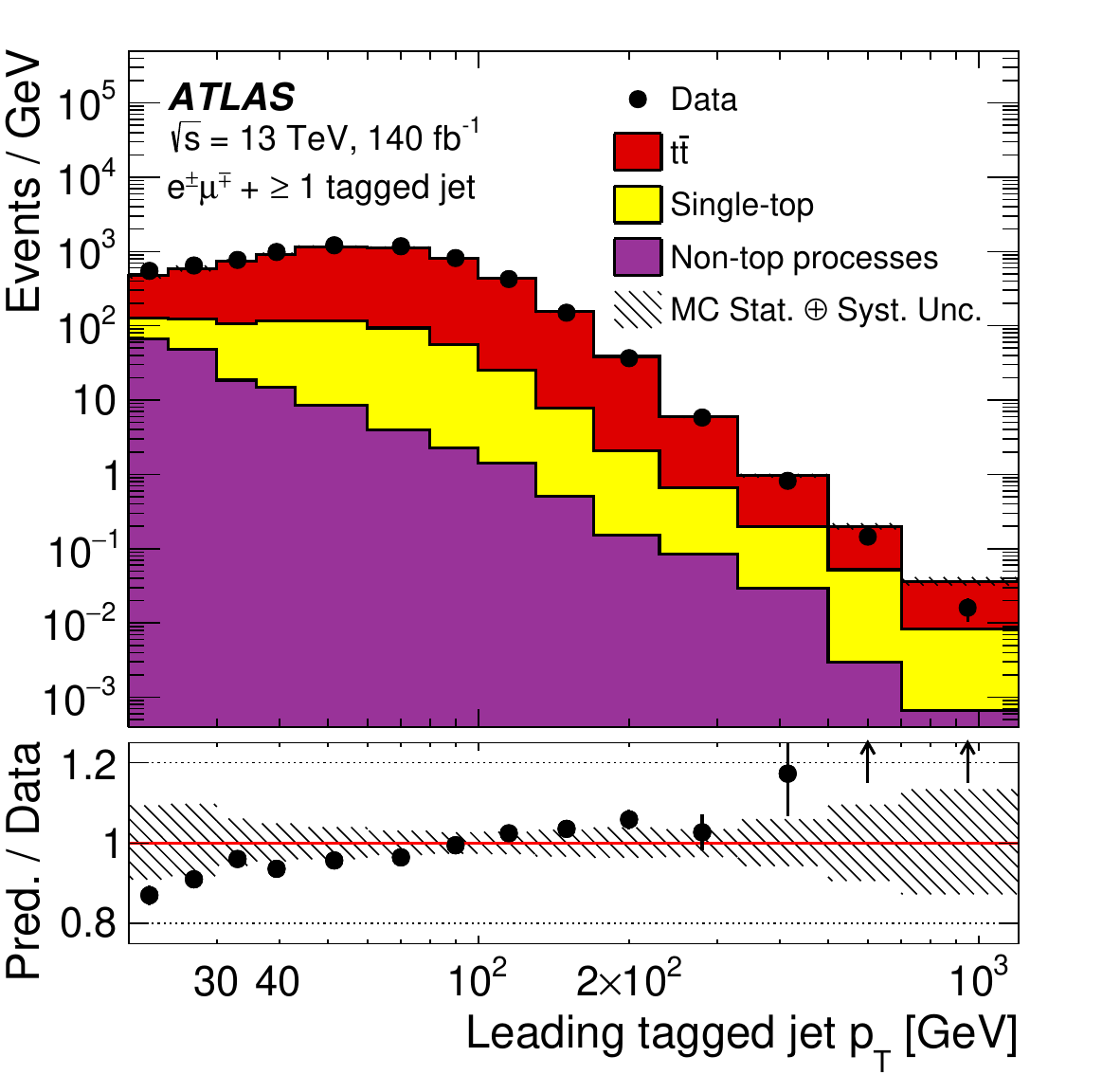}
\caption{}
\end{subfigure}
\begin{subfigure}{0.49\textwidth}
\includegraphics[width=\textwidth]{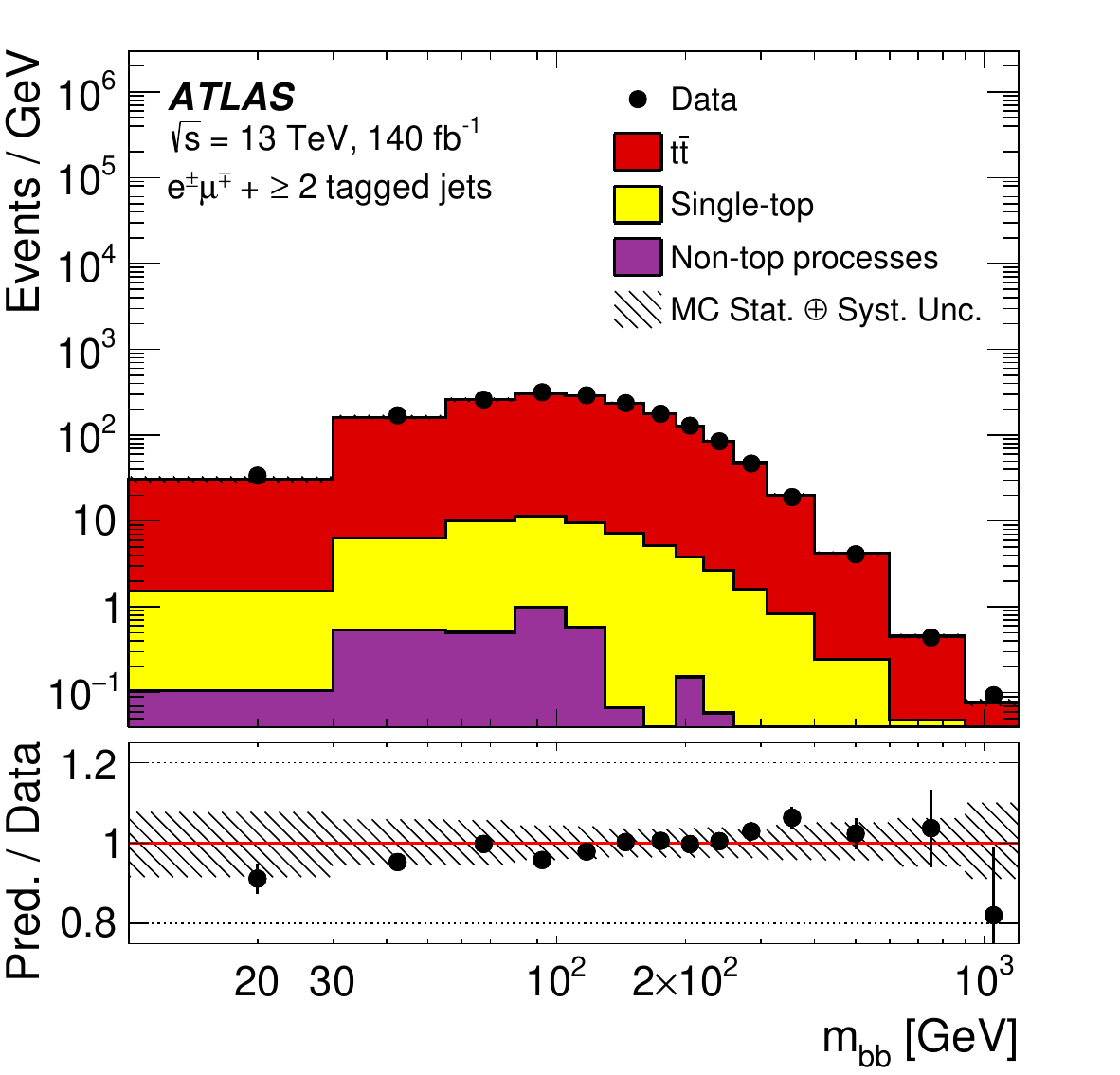}
\caption{}
\end{subfigure}
\caption{
Distribution of events in the $t\bar{t}$ control region as a function of (a) leading tagged jet $p_{\textrm{T}}$  for the 1-tag selection,
and (b) $m_{bb}$  for the 2-tag selection.
The lower panels display the ratio of the predictions of MC simulations for signal plus background to data.
The statistical uncertainty of the data is shown as error bars and the total uncertainty of the prediction as a hatched band (MC theory uncertainties are not included).
}
\label{fig:ttbarCR}
\end{figure}

\subsection{Extraction of \Zjets background}
\label{sec:FlavFit}

The flavour fit consists in a set of maximum-likelihood fits to data used to correct
the simulated shape and rate of \Zjets background components.
The individual fits are based on templates of a jet-flavour sensitive distribution built using events falling in
kinematic regions corresponding, in most of the cases, to the bins of the observable used in the cross-section measurement.
In statistically limited kinematic regions,
events from up to four cross-section bins can be merged to produce the templates.
Such merged bin boundaries are always aligned with those of the cross-section bins.
Each individual fit is done simultaneously in the electron and muon channels,
while fits using each lepton channel independently are checked to yield consistent results.

In the \ZOneTag region, the DL1r $b$-tagging discriminant output of the leading \ftag jet is used as the flavour-sensitive
distribution. This observable is split in four intervals corresponding to certain ranges of $b$-tagging efficiencies,
namely 85--77\% (bin 1 -- loosest \bjet selection, lowest purity), 77--70\% (bin 2), 70--60\% (bin 3), and $<60$\% (bin 4 -- tightest \bjet selection, highest purity).
Separate templates are built for simulated \Zoneb, \Zonec and \Zljets events.
All non-$Z+$jets backgrounds
are combined to build a single template.

In the \ZTwoTag region, the  previously described four intervals of the DL1r discriminant outputs of the leading
and the sub-leading \ftag jets are combined to give a new ten-bin distribution used for fitting.\footnote{
The first four bins include events with one jet in the 85--77\% efficiency range and the other in one of
the four bins as ordered above; bins 5--7 include those with one jet in the 77--70\% range and the other
in one of the three ranges with efficiency below 77\%; bins 8 and 9 correspond to one jet in the 70--60\% range
and the other in the same or $<60$\% range; finally, in bin 10 both jets are in the $<60$\% efficiency range.
}
Templates are built separately for \Ztwob, \Zexoneb, \Zonec and \Zljets events, while all non-$Z+$jets backgrounds
are again combined in a single template.

For all the fits, the normalisation of each \Zjets template is left unconstrained, while
non-$Z+$jets backgrounds are fixed to their MC estimate (and to the data-driven estimate for the \ttbar\ background).
For each measured observable, the fits result in a set of scale factors for each \Zjets\ component, which are the ratios
of their post-fit and pre-fit yields in each bin.
Figure~\ref{fig:Quantiles} shows two examples of flavour-sensitive distributions after the fit in \ZOneTag and \ZTwoTag regions.

\begin{figure}[!tb]
\begin{subfigure}{0.49\textwidth}
\includegraphics[width=\textwidth]{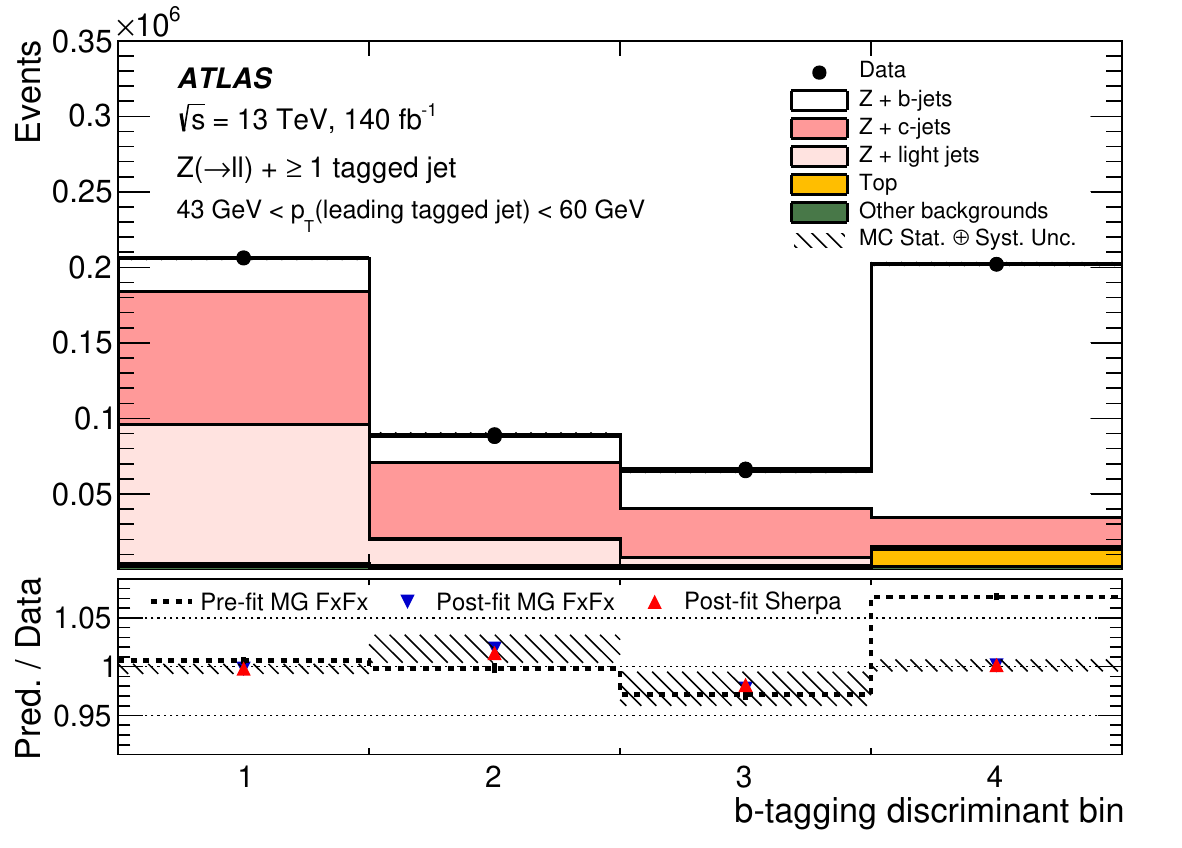}
\caption{}
\end{subfigure}
\begin{subfigure}{0.49\textwidth}
\includegraphics[width=\textwidth]{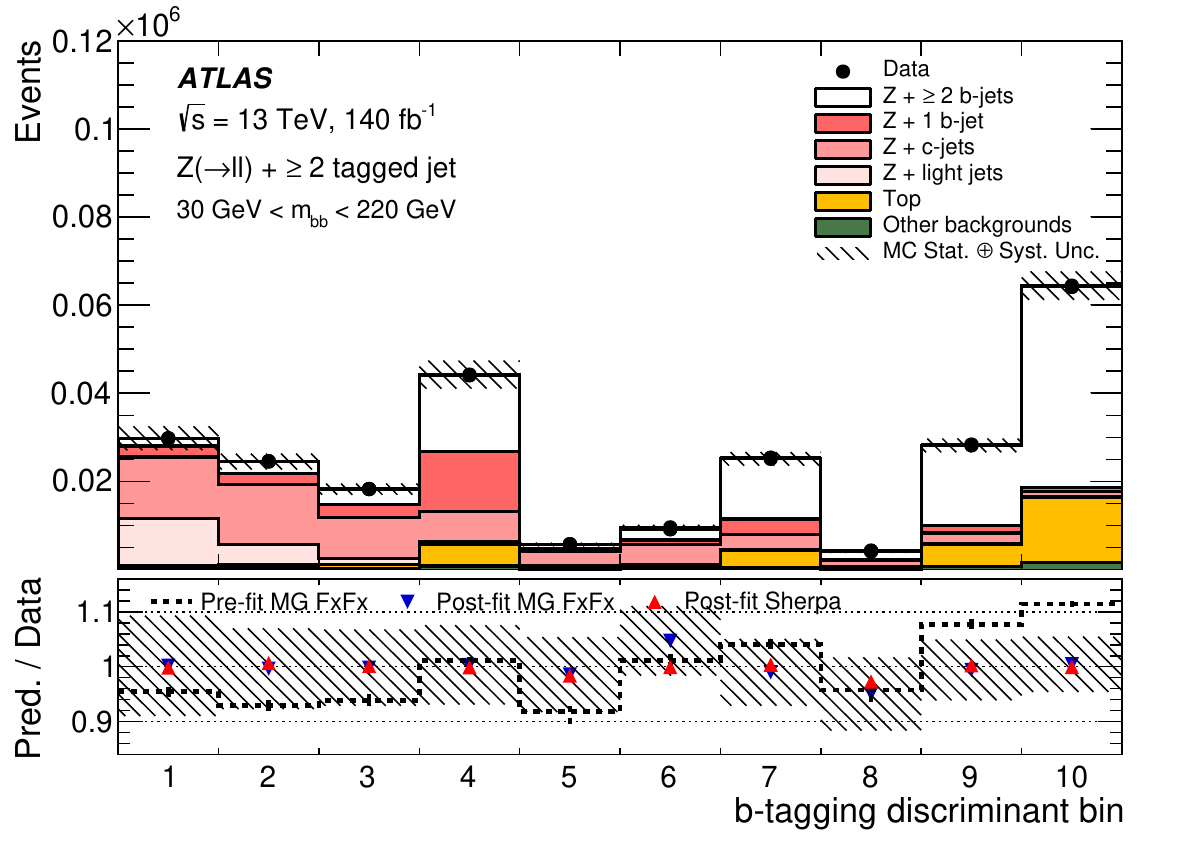}
\caption{}
\end{subfigure}
\caption{
Flavour-sensitive distributions for (a) the 1-tag region fit in the interval  $[43,60]$~\GeV  of $p_{\mathrm{T,}b}^0$ and
(b) the 2-tag region fit in the interval $[30,220]$~\GeV of $m_{bb}$. The main panel compares
data and post-fit distribution. The bottom panels show the ratio of the pre-fit and post-fit distributions to data when using \textsc{MGaMC+Py8 FxFx} and
\textsc{Sherpa} for $Z+$jets simulated samples.
The total uncertainty of the \textsc{MGaMC+Py8 FxFx} post-fit distribution is shown as a hatched band.
}
\label{fig:Quantiles}
\end{figure}

In order to perform the background subtraction from data, the initial MC-based templates of the \Zjets background components, i.e. only \Zljets and \Zcjets for \Zoneb measurement,
same and \Zexoneb for \Ztwob measurement, or only \Zljets and \Zbjets for \Zonec measurement,
are modified using the scale factors in each bin of the flavour fit.
Normalisation scale factors obtained for the signal are not applied to the data.
In cases where binning for the flavour fit is coarser than for the cross-section measurements,
the same scale factor is applied to all bins belonging to the
merged one used for the flavour fit.
The effect of detector-level systematic uncertainties due to the object selection
efficiencies and calibrations, discussed in \secRef{sec:uncertainties}, are assessed by repeating the fits with the templates varied according to each of the
uncertainties. The fit is also repeated for each of the uncertainties affecting the \ttbar\ background and
theory uncertainties of \Zjets and other simulated processes.
An additional systematic uncertainty affecting the \Zjets background shape is
derived by taking the difference between the post-fit \Zjets background evaluated using \mgfxfx\ and
\sherpa\ simulated samples.

\subsection{Estimation of multijet background contribution}
\label{sec:MultiJetBackground}

Background contributions from multijet events is studied separately in the electron and muon channels using a data-driven technique.
Control regions enriched with such events are defined to derive the expected shapes of this background.
These control region definitions are similar to those of the signal regions in Table~\ref{tab:ZBJetSelection} and only lepton selection requirements are different.
In the muon channel, the multijet-enriched control region is characterized by removing the muon isolation criteria
and requiring both muon candidates to have the same charge.
In the electron channel, the identification and impact parameter cuts for electrons are removed, isolation requirement is inverted, and electrons
are required to have the same charge.

In both lepton channels, contributions from non-multijet sources in the control regions are estimated from simulation
and subtracted from the data, with the remaining distributions used as shape templates.
A simultaneous fit of the \mll\ distribution to data in the signal and multijet control region is performed
within the \mbox{40~\GeV $<\mll<$ 160~\GeV} window, in the \ZOneTag and \ZTwoTag regions separately. The normalisation of the \Zjets processes and of the multijet background templates
is left unconstrained in the fit, while the normalisation of the other processes is fixed.

Systematic uncertainties in the multijet contribution are assessed by
excluding the \Zbos\ peak from the fit, performing the fit in the regions without requiring the jets to satisfy flavour tagging
selection (with subsequent extrapolation to the signal regions using normalisation factors equal to the fraction of events in the multijet control region that satisfy the 1-tag and 2-tag requirements) and by allowing the processes other than \Zjets
to be varied independently in the fit.
All these fit variations still yield multijet background estimates consistent with zero.
This background is therefore neglected in the analysis.


%

\section{Kinematic distributions}
\label{sec:distribution}
After requiring the event selection described in Section~\ref{sec:selection}, the measured and expected distributions are compared at the detector level. Pre-fit distributions are used for the signal samples. The \Zjets\ background components are scaled using the respective normalization factors derived from the flavour fit.

Figure~\ref{fig:DataMC} shows the distributions of events in the \ZOneTag signal region as a function of  leading tagged jet \pT (a) and of  leading tagged jet \xF (b), when considering \Zoneb and \Zonec as signal, respectively. The distribution of the events in the \ZTwoTag signal region is presented as a function of \mbb\ (c), when considering \Ztwob as signal.
In each distribution, the \Zjets processes are modelled with \mgfxfx, and \SHERPA is presented in comparison. The uncertainty bands include the uncertainties on the statistics of the simulated sample, on the event-selection (omitting the common luminosity uncertainty) and on the background, as described in Section~\ref{sec:uncertainties}. Theory uncertainties on the generators are not included.
The agreement between the data and the sum of the estimated signal and backgrounds is generally within 10\% for the measurements
in the \ZOneTag region and within 20\% for the \Ztwob measurement (where the signal purity is higher).

\begin{figure}[!tb]
\begin{center}
\begin{subfigure}{0.49\textwidth}
\includegraphics[width=\textwidth]{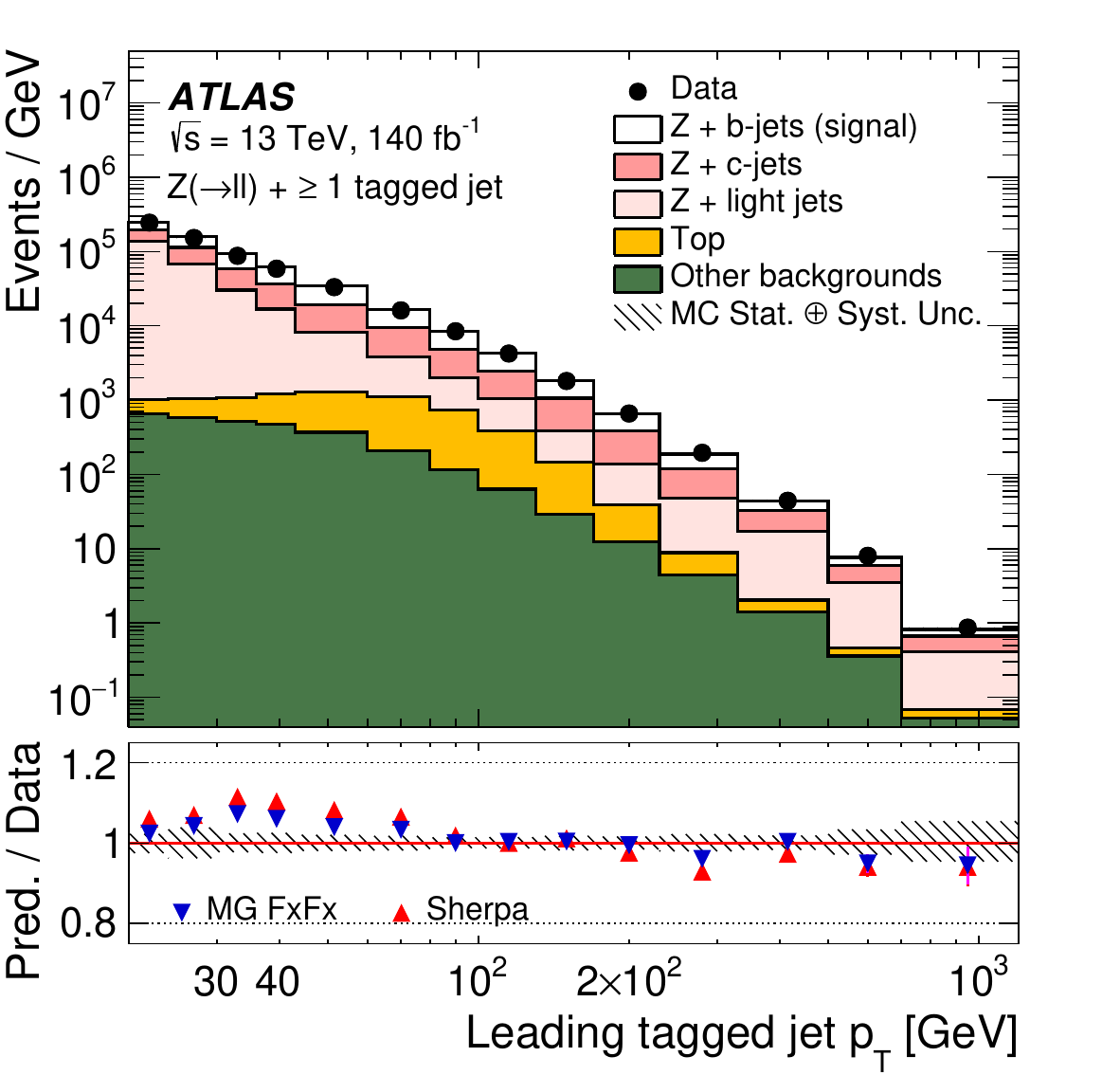}
\caption{}
\end{subfigure}
\begin{subfigure}{0.49\textwidth}
\includegraphics[width=\textwidth]{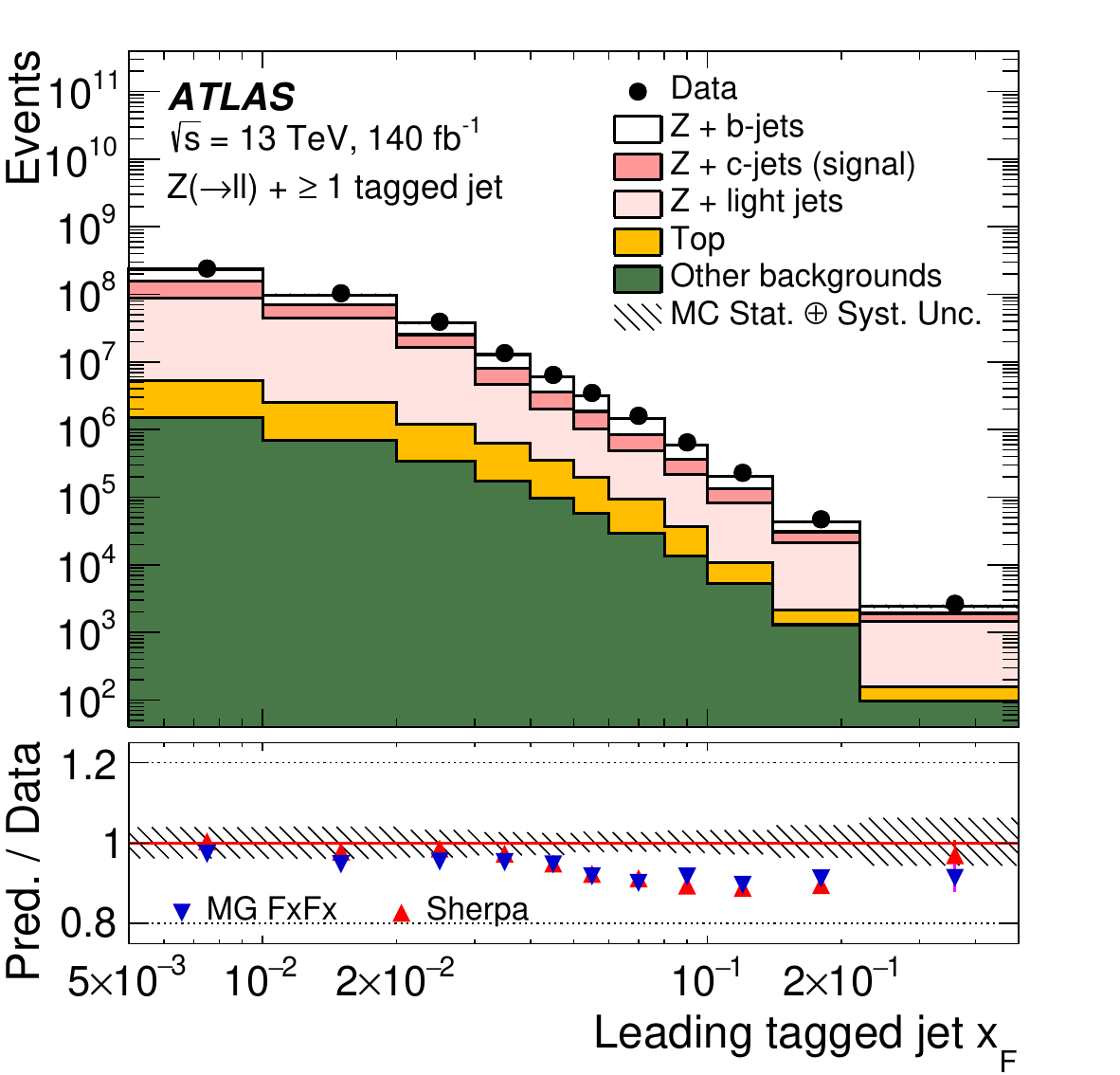}
\caption{}
\end{subfigure}
\begin{subfigure}{0.49\textwidth}
\includegraphics[width=\textwidth]{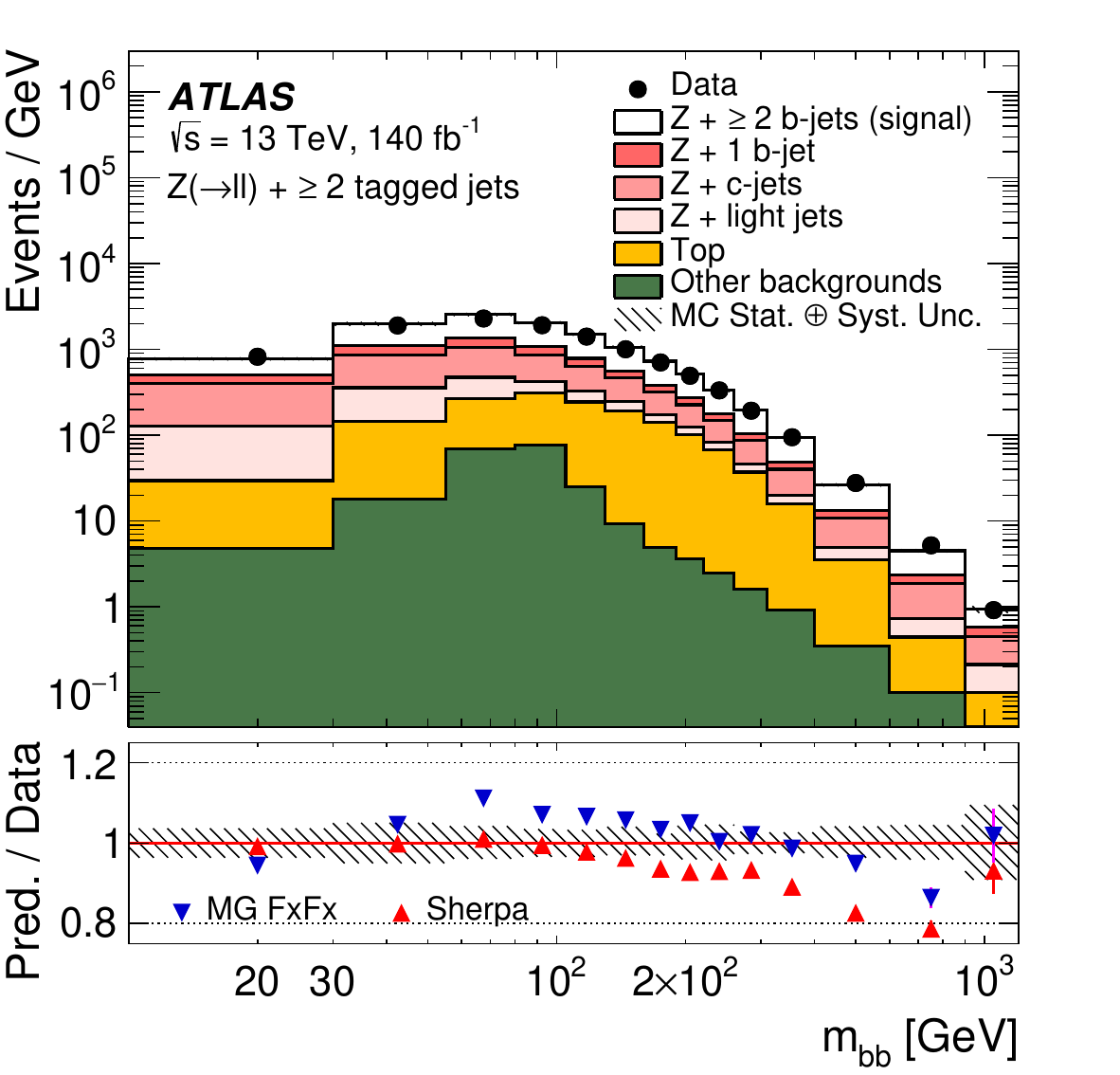}
\caption{}
\end{subfigure}
\end{center}
\caption{
Distribution of events passing the event selection as a function of (a) leading tagged jet $p_{\textrm{T}}$ for $Z\,+ \ge 1~b$-jet as signal and (b) leading tagged jet $x_F$ for $Z\, + \ge 1~c$-jet as signal in the 1-tag region and (c) invariant mass of the leading and sub-leading tagged jets for $Z\, + \ge 2~b$-jets as signal in the 2-tag region. The lower panels display the ratio of the predictions for signal plus background to data using either \textsc{MGaMC+Py8 FxFx} or \textsc{Sherpa 2.2.11} as the signal simulation. The statistical uncertainty of the data is shown as error bars and the total uncertainty of the prediction as a hatched band (MC theory uncertainties are not included).
}
\label{fig:DataMC}

\end{figure}

The total numbers of selected events in data and in predictions are presented in Table~\ref{tab:DLP_Yields}, separately for \Zoneb, \Zonec and \Ztwob measurements, together with the prediction of signal and background process, expressed as a fraction of the total number of predicted events.

\begin{table}[!tb]
\caption{The expected size of the signal and backgrounds, expressed as a fraction of the total number of predicted events for 1-tag and 2-tag signal regions, when considering (a) $Z\,+ \ge 1~b$-jet, (b) $Z\, + \ge 2~b$-jets  or (c) $Z\, + \ge 1~c$-jet as signal. The signal and $Z+$jets background predictions are from the \textsc{MGaMC+Py8 FxFx}  generator, with the $Z+$jets background estimate obtained after applying the normalization scale factors obtained from the flavour fit. The total numbers of predicted and observed events are also shown. The uncertainty in the total predicted number of events is statistical only.
}
\begin{subfigure}{0.325\textwidth}
\footnotesize
\centering

\begin{tabular}{lc}
\toprule
\multicolumn{2}{c}{\ZOneTag region} \\
\midrule
\multicolumn{2}{c}{Signal Z + $\geq$ 1 \bjet } \\
$Z$ + $b$, $Z$ + $bb$			& 34\%\\
\midrule
\multicolumn{2}{c}{Backgrounds} \\
$Z$ + $c$    							& 29\% 	\\
$Z$ + $l$    							& 35\% 	\\
& \\
Top          								& 2\% 	\\
Others       							& 1\% 	\\
\midrule
Total predicted  					& 4\,294\,900 $\pm$ 2100  \\
Data             							& 4\,145\,168	\\
\bottomrule
\end{tabular}

%
%
%
%
%
%
%
%
%
%
%
%
%
%
%
%
%
%
%
%


\caption{}
\end{subfigure}
\begin{subfigure}{0.325\textwidth}
\footnotesize
\centering

\begin{tabular}{lc}
\toprule
\multicolumn{2}{c}{\ZTwoTag region} \\
\midrule
\multicolumn{2}{c}{Signal Z + $\geq$ 2 \bjets } \\
$Z$ + $bb$			& 46\%\\
\midrule
\multicolumn{2}{c}{Backgrounds} \\
$Z$ + $b$    							& 11\% 	\\
$Z$ + $c$    							& 23\% 	\\
$Z$ + $l$    							& 7\% 	\\
Top          								& 12\% 	\\
Others       							& 2\% 	\\
\midrule
Total predicted  					& 325\,300 $\pm$ 600 \\
Data             							& 309\,199	\\
\bottomrule
\end{tabular}

%
%
%
%
%
%
%
%
%
%
%
%
%
%
%
%
%
%
%
%


\caption{}
\end{subfigure}
\begin{subfigure}{0.325\textwidth}
\footnotesize
\centering

\begin{tabular}{lc}
\toprule
\multicolumn{2}{c}{\ZOneTag region} \\
\midrule
\multicolumn{2}{c}{Signal Z + $\geq$ 1 \cjet } \\
$Z$ + $c$								& 28\%	\\
\midrule
\multicolumn{2}{c}{Backgrounds} \\
$Z$ + $b$, $Z$ + $bb$  							& 33\% 	\\
$Z$ + $l$    							& 37\% 	\\
& \\
Top          								& 2\% 	\\
Others       							& 1\% 	\\
\midrule
Total predicted  					& 3\,994\,400 $\pm$ 2000 \\
Data             							& 4\,145\,168	\\
\bottomrule
\end{tabular}

%
%
%
%
%
%
%
%
%
%
%
%
%
%
%
%
%
%
%
%


\caption{}
\end{subfigure}
\label{tab:DLP_Yields}
\end{table}


%

\section{Correction to particle level}
\label{sec:unfolding}

The signal event yields are determined by subtracting the estimated background contributions from the data. The resulting distributions are corrected for detector-level effects, to the fiducial phase space at particle level, as defined in Table~\ref{tab:fiducial_ps}. Particle-level objects are selected with requirements close to the reconstruction level selection described in Section~\ref{sec:selection} to limit the dependence of the measurement on theoretical predictions. In this definition, the lepton kinematic variables are computed using final-state leptons from the \Zbos\ decay. Radiated photons within a cone of $\Delta R = 0.1$ around the direction of a final-state lepton are added to the lepton, and the sum is referred to as the `dressed' lepton. Particle-level jets are identified by applying the anti-$k_t$  algorithm with $R = 0.4$ to all final-state particles with a lifetime longer than 30~ps, excluding the dressed \Zbos\ decay products.
Particle-level jets are classified as \bjets or \cjets following the same logic as described for reconstructed jets
in Section~\ref{sec:selection}.
The correction to particle-level accounts for selection efficiencies, resolution effects and differences between the fiducial and the detector-level phase spaces.

\begin{table}[!h]
\begin{center}
\caption{Fiducial regions of the $Z\,+ \ge 1~b$-jet, $Z\, + \ge 2~b$-jets and $Z\, + \ge 1~c$-jet measurements at particle level. The assignment of a $b$-jet or a $c$-jet is explained in \secRef{sec:selection}.}
\begin{tabular}{c|l}
\toprule
Object Selection     			& Acceptance cuts  \\
\midrule
\multirow{2}{*}{Lepton} 		&  $\pt>27$~GeV, $|\eta|<2.5$ \\
&  2 same flavour and opposite charge, $76~\GeV < m_{\ell\ell} < 106~\GeV$ \\

$b$-jet   								&  $\pt>20$~GeV, $|y|<2.5$, $\Delta R(b\text{-jet},\ell)>0.4$\\
$c$-jet   								&  $\pt>20$~GeV, $|y|<2.5$, $\Delta R(c\text{-jet},\ell)>0.4$\\
\midrule
Event Selection  & Acceptance cuts  \\
\midrule
$Z\,+ \ge 1~b$-jet								& $Z\,+ \ge 1~b$-jet and a $b$-jet is the leading heavy-flavour jet\\
$Z\, + \ge 2~b$-jets 								& $Z\, + \ge 2~b$-jets and a $b$-jet is the leading heavy-flavour jets\\
$Z\,+ \ge 1~c$-jet 								& $Z\,+ \ge 1~c$-jet and a $c$-jet is the leading heavy-flavour jet\\
\midrule
Rapidity regions & Acceptance cuts \\
\midrule
Central rapidity & $Z$ boson rapidity $|y(Z)| < 1.2$ \\
Forward rapidity & $Z$ boson rapidity $|y(Z)| \ge 1.2$ \\
\bottomrule
\end{tabular}
\label{tab:fiducial_ps}
\end{center}
\end{table}

Differential distributions are corrected to the particle level by using an iterative Bayesian unfolding~\cite{DAgostini} with two iterations.
Simulated signal events, passing the selection defined in Section~\ref{sec:selection}, are used to generate a response matrix for each distribution. The matrix is filled with the events that pass both the detector-level and particle-level selections and accounts for bin-to-bin migration effects between the detector-level and particle-level distributions.
Figure~\ref{fig:Matrices} shows two examples of these response matrices.
In the first iteration of the Bayesian unfolding, the particle-level prediction is used as the initial prior. From the second iteration, the prior is given by the unfolded distribution of the previous iteration and the unfolding matrix is derived on the basis of the Bayes' theorem, from the response matrix and the prior. The use of more than two iterations has not been found to change significantly the final results, but it would have increased the statistical uncertainty.

\begin{figure}[h!]
\begin{center}
\begin{subfigure}{0.49\textwidth}
\includegraphics[width=\textwidth]{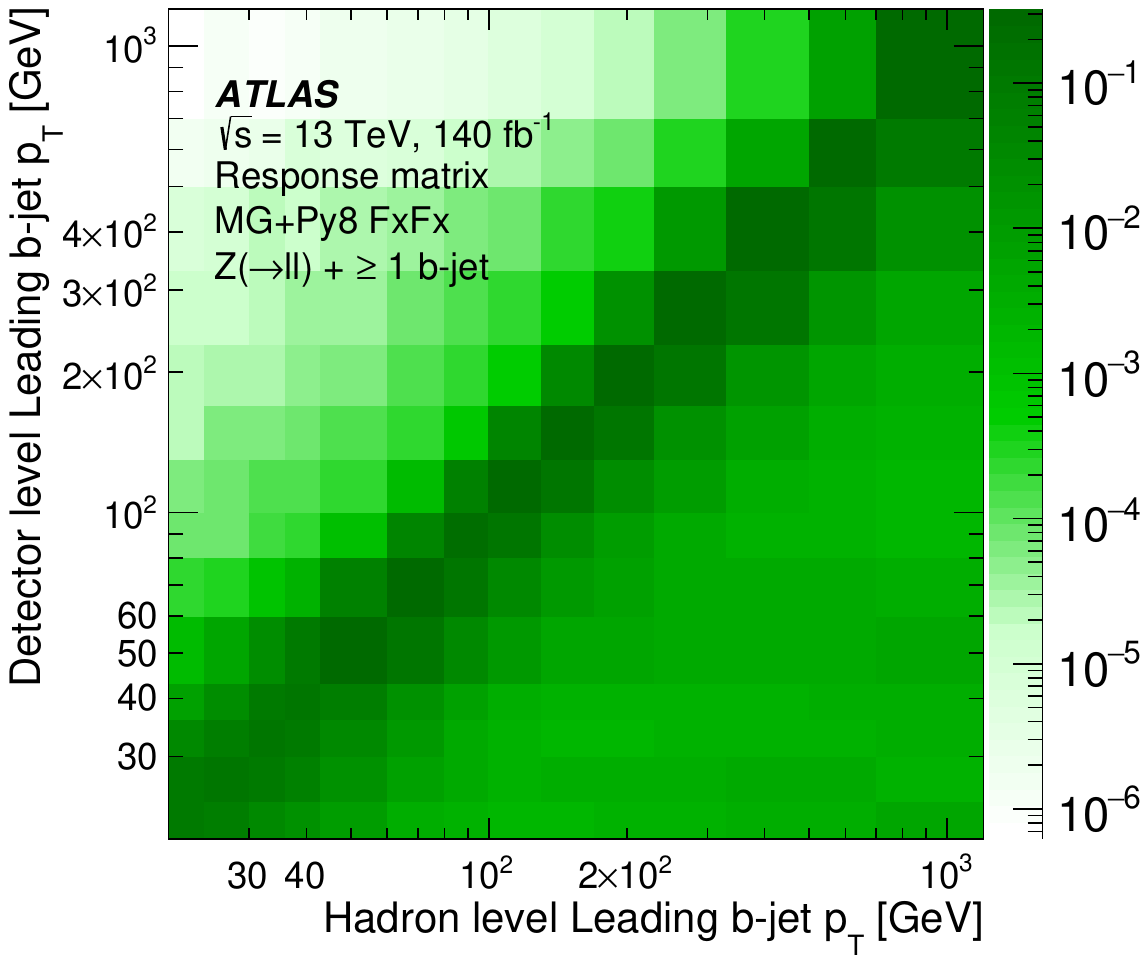}
\caption{}
\end{subfigure}
\begin{subfigure}{0.49\textwidth}
\includegraphics[width=\textwidth]{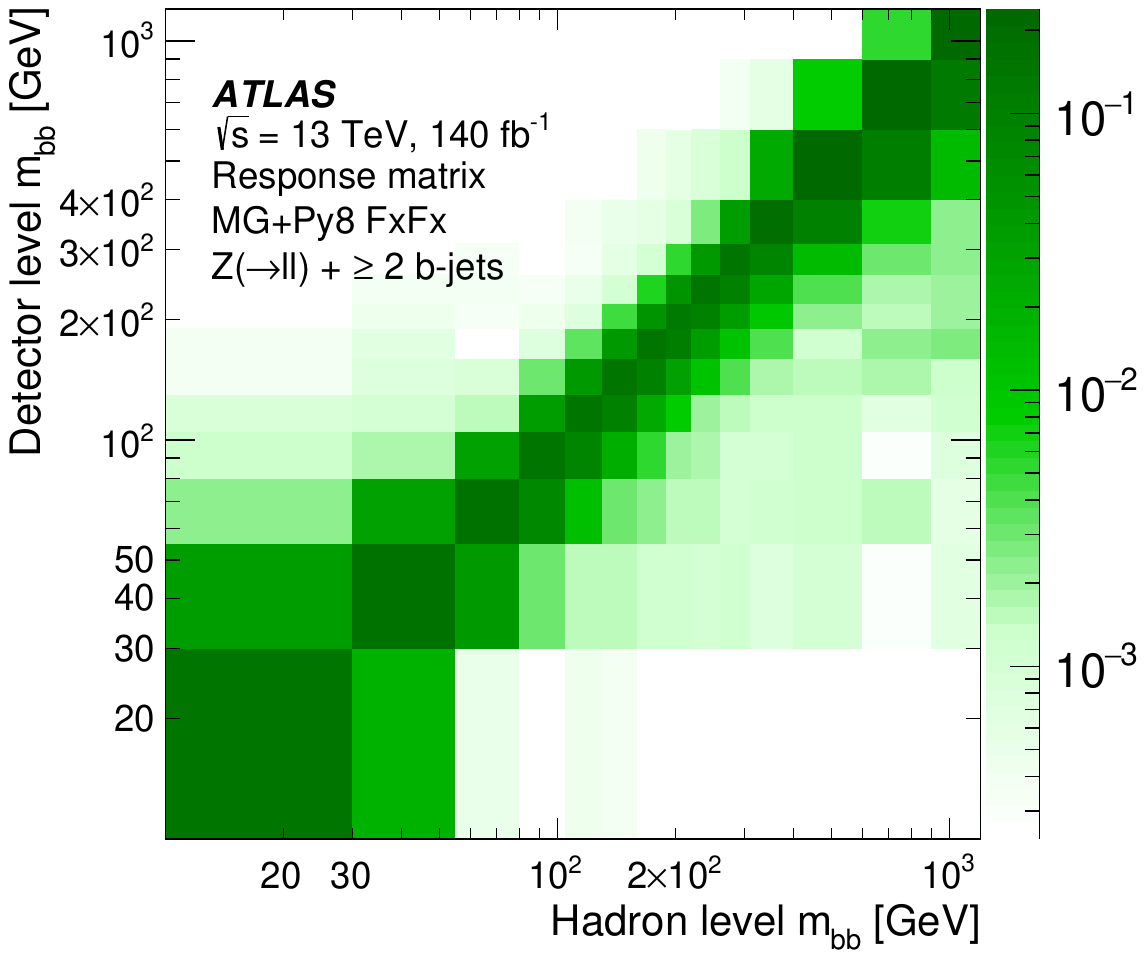}
\caption{}
\end{subfigure}
\end{center}
\caption{
Response matrix of (a) leading $b$-jet  $p_{\textrm{T}}$ for $Z + \ge 1 b$-jet events and (b) $m_{bb}$ for $Z + \ge 2 b$-jets events.
The matrix is created from the \textsc{MGaMC+Py8 FxFx} sample and used for the nominal unfolding of the data.
The sum of the entries in each column is normalised to the reconstruction efficiency in corresponding bin.
}
\label{fig:Matrices}
\end{figure}

The pre-fit distributions of the \mgfxfx signal samples are used to perform the unfolding procedure. The background-subtracted data are corrected for the expected fraction of events that pass the detector-level selection, but not the particle-level one (unmatched events), before entering the iterative unfolding.
To obtain the cross-sections, the unfolded event yields are divided by the integrated luminosity of the data sample and by the bin width in each bin of each differential distribution.
The differential cross-section measurement of a given observable in the $i$-th bin is given by:
\[
\sigma_i = \dfrac{1}{\epsilon _i L} \sum\limits_j U_{ij} f_j   N_j^{\textrm{bsD}}\textrm{,}
\]
where $L$ is the integrated luminosity, $\epsilon_i$ is the reconstruction efficiency in the $i$-th bin, $N_j^{\textrm{bsD}}$ is the number of background-subtracted data events in the reconstructed $j$-th bin, $f_j$ is the factor that corrects for unmatched events in the $j$-th bin, and $U_{ij}$ is the element $(i,j)$ of the unfolding matrix calculated after two iterations, using the updated prior from the first iteration and the response matrix.
The efficiency is defined as a fraction of simulated events in $i$-th particle-level bin that pass the detector-level selection,
and the $f_j$ is a fraction of simulated events reconstructed in $j$-th bin that also pass the fiducial particle-level selection.

The measurement of  \RptZ, as defined in Table~\ref{tab:fiducial_ps}, is performed as follows:
At reconstruction level, the background subtraction is performed separately for events passing the central and forward rapidity selection, as defined in Table~\ref{tab:ZBJetSelection}.
Resulting distributions of \Zbos \pT are then unfolded simultaneously in the full fiducial phase space, to
obtain the corresponding particle-level distributions, accounting for migrations between the rapidity
regions via the unfolding matrix (in addition to migrations between the \Zbos \pT bins).
Finally, the unfolded distribution for the forward region is divided by the one for the central region.
In this way all systematic uncertainties are treated coherently between the rapidity regions, allowing a significant cancellation of the
correlated ones.

The measurement of the inclusive fiducial cross-section for \Zoneb, \Ztwob and \Zonec processes is
obtained by applying a particle-level correction to the number of events in data passing the selection in \secRef{sec:selection}, after background subtraction\footnote{
For the purpose of flavour fit and data-driven \ttbar background evaluation procedure,
all events are treated as belonging to a single bin.}.
The  correction, which is applied as a divisor of the background-subtracted data, is derived from the ratio of the total number of reconstructed events in the detector-level phase space to the number of particle-level events in the fiducial phase space.
Integration of the unfolded differential cross-sections yield consistent results, but the described procedure
is adopted as the one independent of the choice of binning for the flavour fit and the unfolding.

Since the electron and muon decay channels are combined to increase the precision of the signal fits to data, the corrections and response matrices are made using electron and muon signal samples to obtain combined particle-level yields.
To validate this procedure, the analysis is performed for each of the two lepton channels separately.
The results obtained from the individual channels are compatible within $1\sigma$ for the inclusive fiducial cross-section of \Zoneb and \Zonec production and within $2\sigma$ for the \Ztwob cross-section.
This comparison uses only the sum in quadrature of the statistical and uncorrelated systematic uncertainties.
The differential cross-section measurements in the two channels also agree over the full range of each distribution.


%

\section{Uncertainties in the cross-section measurement}
\label{sec:uncertainties}

Uncertainties in the \Zoneb, \Ztwob\ and \Zonec\ cross-section measurements arise from systematic effects related to detector-level selection, background determination and unfolding method and from the statistical uncertainty of the analysed data sample. Uncertainties are considered correlated, when appropriate, between lepton channels, signal and background processes and over the observables.

\subsection{Detector-level systematic uncertainties}
Detector-level systematic uncertainties are defined to be the uncertainties in the selection of the physics objects entering the measurement and in the luminosity. They are derived for each observable by propagating shifts from each systematic source through both the unfolding inputs (response matrices, reconstruction efficiency and the unmatched events correction) and the subtracted background into the unfolded data, after the flavour fit.

Systematic uncertainties on the leptons are related to the trigger, reconstruction, identification and isolation criteria adopted in the selection
of electrons and muons~\cite{EGAM-2018-01, MUON-2018-03}.
Variations in the electron energy scale and resolution are taken into account, as are those related to the muon momentum scale, the inner-detector and muon spectrometer resolution, and the sagitta-bias corrections.
These uncertainties are specific for each leptonic final state and therefore uncorrelated between the two lepton channels. The contribution of the lepton uncertainties is less than 1\% of the final measurements.

The uncertainties associated to the reconstructed jets take into account corrections to the energy scale (JES) and resolution (JER)~\cite{PERF-2016-04}.
They are estimated by scaling and smearing the jet four-momentum in the simulation by the associated uncertainties in the calibration procedure.
The JVT selection efficiency is also considered. Jet systematic uncertainties are among the largest uncertainties in the measured cross-sections, with an average contribution of 5\% which increases in specific regions of phase space (i.e. high \mbb).

The systematic uncertainty on \met\ accounts for the energy scale and resolution of the soft hadronic activity reconstructed in the event~\cite{ATLAS:2024cmj} and contribute to less than 1\%.

The flavour tagging uncertainties are derived from the calibration of the DL1r tagger using
data control samples enriched in $b$~\cite{FTAG-2018-01}, $c$~\cite{FTAG-2020-08}, or light-jets~\cite{FTAG-2019-02}, up to a jet \pt of several hundreds GeV.
Any difference in the flavour-tagging performance measured in data events and MC samples is used to correct the flavour-tagging efficiency in the MC\@
as a function of the jet flavour, the different $b$-tagging discriminant output thresholds, and of the jet \pt.
In the case of \bjets, correction factors are close to unity  and their uncertainties, described by a set of 45 independent parameters, are as low as 1\% for jet \pt\ of about 60~\GeV, but reach 7\% for jet \pt\ of about 20~\GeV\ and up to 3\% at 400~\GeV.
In the case of \cjets, correction factors range from about 1 to about 1.3, their uncertainties are described by a set of 20 independent parameters, and are about 3-4\% in the bulk of the phase space, but reach up to 15\% for low jet \pt\ and for the largest values of the $b$-tagging discriminant output.
In the case of light-flavour jets, correction factors for light-jets  range from about 1 to about 1.3, with uncertainties described by a set of 20 independent parameters and ranging from 10\% to 20\%.
The uncertainties in the flavour-tagging calibration are extrapolated to  high \pT on the basis of MC simulations.

The flavour tagging systematic is the second largest uncertainty in \Zoneb and \Ztwob inclusive cross-sections, with average contributions of respectively 3.6\% and 5.7\%, and it becomes the largest uncertainty in the \Zonec measurement with an average contribution of 10.3\% to the final precision.

The uncertainty of the imperfect modelling of the pile-up effects is assessed by varying the average number of pile-up interactions. It contributes to less than 1\% to the cross-section measurements.
The uncertainty in the combined 2015–2018 integrated luminosity is 0.83\%~\cite{DAPR-2021-01}, obtained using the LUCID-2 detector for the primary luminosity measurements~\cite{LUCID2}.

\subsection{Background systematic uncertainties}
The uncertainty from each background source is determined by applying shifts to the subtracted background contributions and to the nominal unfolding inputs.

The uncertainty related to the determination of the \Zjets background ranges from 0.6\% to 1.6\% in the measured inclusive cross-sections. It is determined as an envelope of two separate but related variations. The first accounts for the differences between the post-fit \Zjets backgrounds evaluated using \mgfxfx or \SHERPA, as described in Section~\ref{sec:FlavFit}. The second source accounts for the \mgfxfx theoretical uncertainties, which are determined by varying independently the QCD renormalisation ($\mu_R$) and factorisation ($\mu_F$) scales by a factor 0.5 and 2 with an additional constraint of $0.5 < \mu_R/\mu_F < 2$. For each of these scale variations the flavour fit is repeated and an envelope of the post-fit result is used for the error estimate.
Effects due to PDF and $\alpha_{s}$ uncertainties have a negligible impact and are not propagated to the final measurements.

The systematic uncertainty in the data-driven $t\bar{t}$ background accounts for the extrapolation from the $e^{\pm}\mu^{\mp}$ control region to the signal region, as explained in Section~\ref{sec:DataDrivenTTbar}. The contribution of this uncertainty is less than 1\% in the inclusive cross-sections.

The uncertainty in the other backgrounds is given by the sum in quadrature of the uncertainties in the MC-modelled diboson, single-top, $ZH$ and $Z\rightarrow\tau\tau$ backgrounds. Diboson uncertainty is determined by varying independently the QCD scales $\mu_R$ and $\mu_F$ by a factor 0.5 and 2 with an additional constraint of $0.5 < \mu_R/\mu_F < 2$ and by taking the envelope of the predicted variations as error estimate. For the smaller single-top, $ZH$ and $Z\rightarrow\tau\tau$ contaminations, overall normalization uncertainties covering QCD scale, PDF and $\alpha_{s}$ variations, are considered. The contribution to the measured cross-sections is negligible.

\subsection{Unfolding systematic uncertainties}

The uncertainties in the unfolding procedure originate from the statistical uncertainty of \mgfxfx used as nominal MC in the unfolding, the intrinsic bias introduced by the unfolding method and the modelling and theoretical uncertainties of the \mgfxfx signal samples.

The uncertainty due to the limited statistics of \mgfxfx samples is propagated using 100 MC pseudo-experiments: the unfolding inputs are fluctuated independently according to Gaussian distributions and the RMS of the results is taken as error estimate.

The dependence of the Bayesian unfolding regularisation procedure on the choice of the initial prior is determined by reweighting the generator-level distribution of each observable in the \mgfxfx samples to provide a better description of the data at detector level. The modified \mgfxfx samples are then used to emulate data and are unfolded with the nominal matrix.

The signal modelling uncertainty accounts for the mis-modelling of the migrations, the reconstruction efficiency and the unmatched events corrections. It is evaluated by comparing unfolded results when using different MC generators. First, for each measured observable, \SHERPA signal sample is reweighted to match the \mgfxfx particle-level distribution of that observable and used to unfold its distribution in data.
The reweighting is performed to avoid double-counting with the estimate of the effect of the initial prior choice described above.
Second, theoretical uncertainties related to modelling the signal in \mgfxfx are evaluated by the same variations of QCD scales as described above and taking an envelope of the unfolded results using the varied matrices.

The contributions to the unfolding uncertainty are summed in quadrature and represent one of the dominant uncertainties for the \Zoneb, \Ztwob and \Zonec inclusive cross-sections with contributions of respectively 3.3\%, 5.8\% and 5.0\%.

Table~\ref{tab:unc_Zll} summarizes the systematic uncertainties in the measured inclusive fiducial cross-sections for the \Zoneb, \Ztwob and \Zonec production.
Figures~\ref{fig:sys_break_example_Zb}--\ref{fig:sys_break_example_Zc} show, as examples, the breakdown of the systematic uncertainties in the cross-section as a function of
\Zbos \pt for events with at least one \bjet,
\mbb for events with at least two \bjets,
leading \cjet \pT for events with at least one \cjet,
and in the ratio of the cross-sections as functions of \Zbos \pT in forward and central \Zbos rapidity regions.

\begin{table}[!tb]
\begin{center}
\caption{Relative systematic uncertainties in the measured production cross-sections of $Z\,+ \ge 1~b$-jet, $Z\, + \ge 2~b$-jets and $Z\, + \ge 1~c$ events.
The ``Jet'' term includes the JES, JER and JVT uncertainties.
The ``Lepton'' term includes the lepton trigger, efficiency, scale and resolution uncertainties.}

\begin{tabular}{l|c|c|c}\toprule
Source of uncertainty& $Z(\to \ell\ell)~+ \ge1$ \bjet & $Z(\to \ell\ell)~+ \ge2$ \bjets & $Z(\to \ell\ell)~+ \ge1$ \cjet\\
& [$\%$] &[$\%$] \\
\midrule
Flavour tagging                      		& 3.6           & 5.7          & 10.3 \\
Jet                                 				& 2.4          & 4.3          & 6.5\\
Lepton                              			& 0.3           & 0.3          & 0.4\\
$\MET$                              & 0.4           & 0.5          & 0.3\\
\Zjets background & 0.6           & 1.5          & 1.6\\
Top  background                     & 0.1           & 0.3          & $<$0.1\\
Other backgrounds                   & $<$0.1           & 0.2          & 0.1\\
Pile-up                             & 0.6           & 0.6          & 0.2\\
Unfolding                           & 3.3           & 5.8          & 5.0\\
Luminosity                          & 0.8           & 0.9          & 0.7\\
\midrule
Total [\%]                          & 5.6           & 9.4          & 13.2\\
\bottomrule
\end{tabular}


\label{tab:unc_Zll}
\end{center}
\end{table}

\begin{figure}[!tb]
\begin{center}
\begin{subfigure}{0.49\textwidth}
\includegraphics[width=\textwidth]{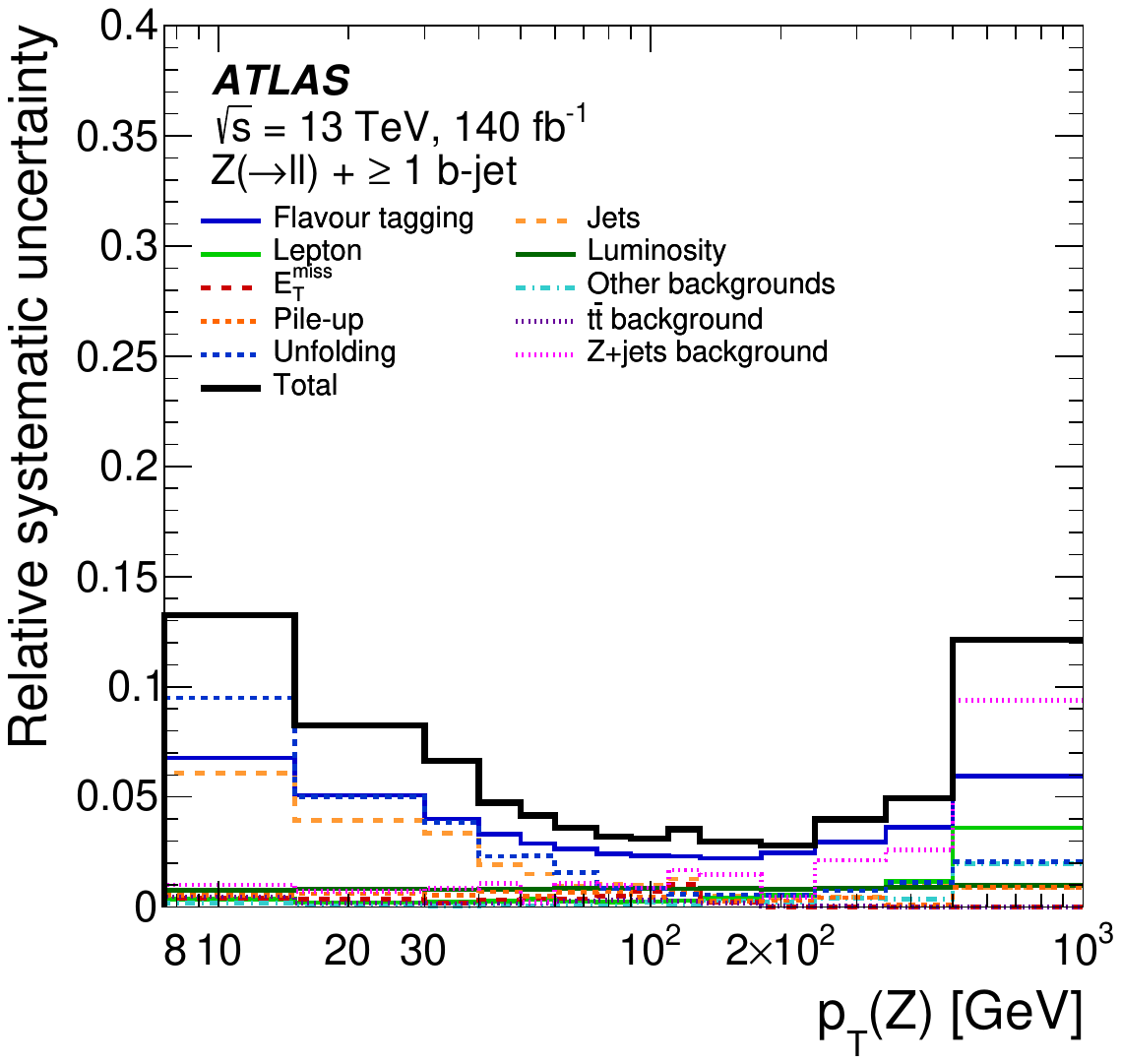}
\caption{}
\end{subfigure}
\begin{subfigure}{0.49\textwidth}
\includegraphics[width=\textwidth]{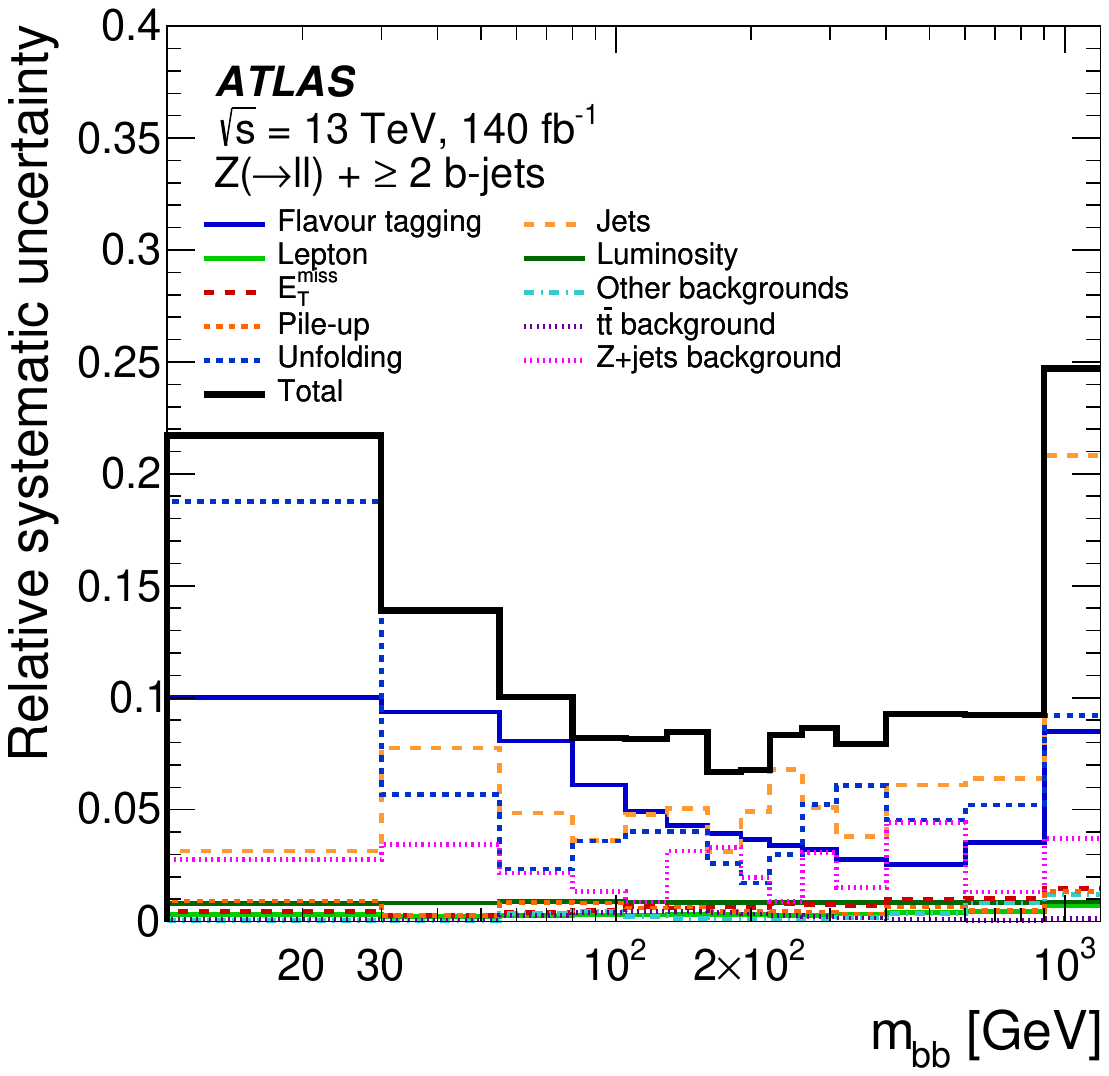}
\caption{}
\end{subfigure}
\end{center}
\caption{
Relative systematic uncertainties in the fiducial cross-section (a) as a function of $p_{\textrm{T}}(Z)$ in events with at least one $b$-jet
and (b) as a function of $m_{bb}$  in events with at least two $b$-jets.
The total uncertainty is shown with solid black line while the different components listed in Table~\ref{tab:unc_Zll} are shown in different line styles and colours.}
\label{fig:sys_break_example_Zb}
\end{figure}

\begin{figure}[!tb]
\begin{center}
\begin{subfigure}{0.49\textwidth}
\includegraphics[width=\textwidth]{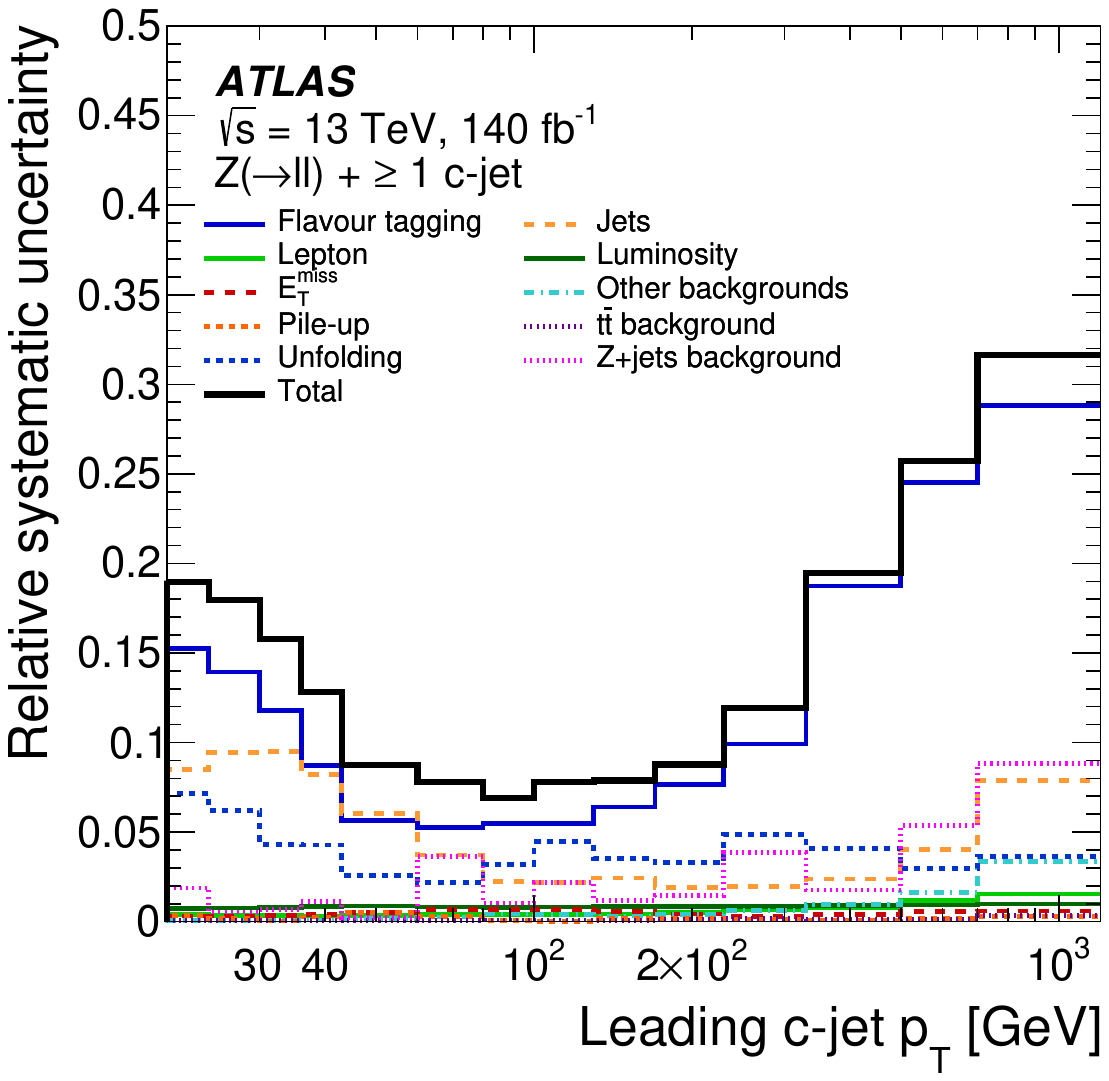}
\caption{}
\end{subfigure}
\begin{subfigure}{0.49\textwidth}
\includegraphics[width=\textwidth]{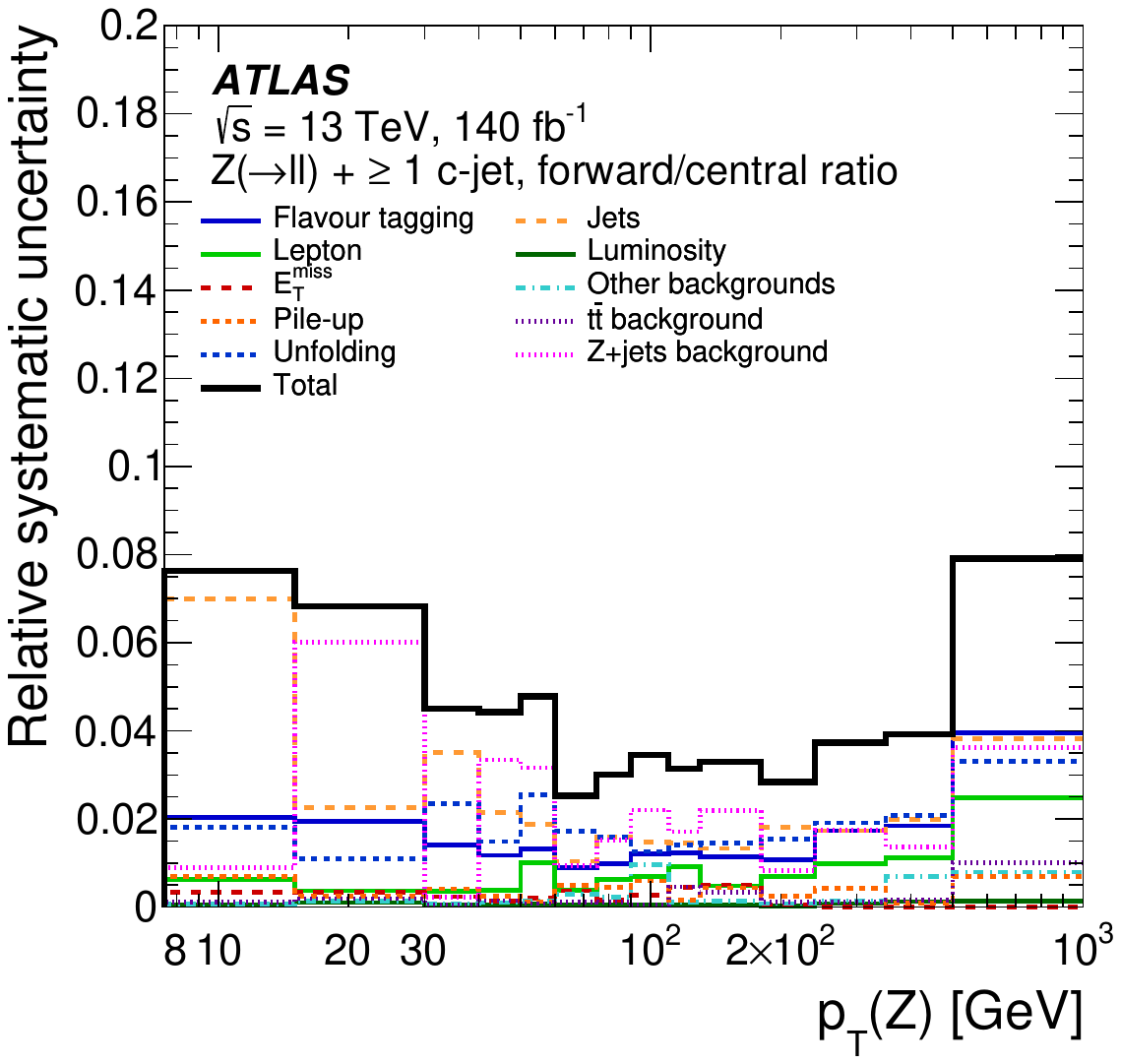}
\caption{}
\end{subfigure}
\end{center}
\caption{
Relative systematic uncertainties in
(a) the fiducial cross-section as a function of leading $c$-jet $p_{\textrm{T}}$  ($p_{\mathrm{T,}c}^0$) and
(b) the ratio of the cross-sections as functions of $Z$ boson $p_{\textrm{T}}$  in forward and central $Z$ boson rapidity regions in events with at least one $c$-jet.
The total uncertainty is shown with solid black line while the different components listed in Table~\ref{tab:unc_Zll} are shown in different line styles and colours.}
\label{fig:sys_break_example_Zc}
\end{figure}

The total systematic uncertainty in the inclusive cross-sections is 5.6\% in \Zoneb\ events, 9.4\% in \Ztwob\ events and 13.2\% in \Zonec events.
In the differential distributions it is less than 5\% in the \Zoneb, except in some bins of \ptZ. In \Ztwob\ and \Zonec\ measurements it is at a level of 10\%--15\%,
except in some bins at the edges of the distributions.

The statistical uncertainty of the data is propagated through the unfolding by using 1000 pseudo-experiments. The flavour fit and the unfolding is repeated for each data replica. The statistical uncertainty in the inclusive cross-sections of \Zoneb, \Ztwob\ and \Zonec\ is 0.2$\%$, 0.4$\%$ and 0.3$\%$ respectively.


%


\section{Results}
\label{sec:results}

The inclusive and differential cross-section measurements for \Zoneb, \Ztwob\ and \Zonec\ are shown
in Figures~\ref{fig:pred_XS_incl_B}--\ref{fig:pred_fiducial1C_xF_and_Ratio}.
The measurements are compared with standard 5FS multi-leg \mgfxfx\ and \SHERPA\ predictions, \mgfxfx\ variations
with several PDFs using different intrinsic charm models, with NLO \mgamc\ predictions of various flavour schemes and
with fixed-order 5FS NLO and NNLO predictions using the flavour-dressing algorithm (see Section~\ref{sec:datasets}).

Theoretical uncertainties of the various samples, computed as described in Section~\ref{sec:datasets}, are shown
in the comparison with data. In this section, all predictions are normalised to their own cross-section to allow an
unbiased comparison among different generators.

\subsection{Inclusive fiducial cross-sections}

The measured inclusive cross-sections in the fiducial phase space for \Zoneb, \Ztwob\ and \Zonec, shown in Figures~\ref{fig:pred_XS_incl_B} and \ref{fig:pred_XS_incl_C},
are $10.49\pm 0.02 (\textrm{stat.})\pm 0.59 (\textrm{syst.})$~pb,
$1.39\pm 0.01 (\textrm{stat.})\pm 0.13 (\textrm{syst.})$~pb,
and $20.9\pm 0.1 (\textrm{stat.})\pm 2.8 (\textrm{syst.})$~pb, respectively.

The 5FS simulations, in general, adequately predict the inclusive cross-sections for both the \Zoneb\ and \Ztwob whereas the 4FS simulation shows an underestimate of about $2\sigma$ for the  \Zoneb\ inclusive
cross-section, while predicting the \Ztwob\ cross-sections accurately. Overall, these results are consistent with the ones presented in the ATLAS measurement on a partial Run~2 data set~\cite{STDM-2017-38},
based on previous generator versions.
All 5FS multi-leg predictions considered here, as well as \mgamcfour, are in agreement with the measured \Zonec\ cross-section.
The \mgamcthree prediction drastically underestimates
the measured cross-section by about 50\%, consistent with the lack of resummation of $\ln(Q^2/m_c^2)$ in the collinear PDF evolution.

\begin{figure}[!tb]
\begin{center}
\begin{subfigure}{0.49\textwidth}
\includegraphics[width=\textwidth]{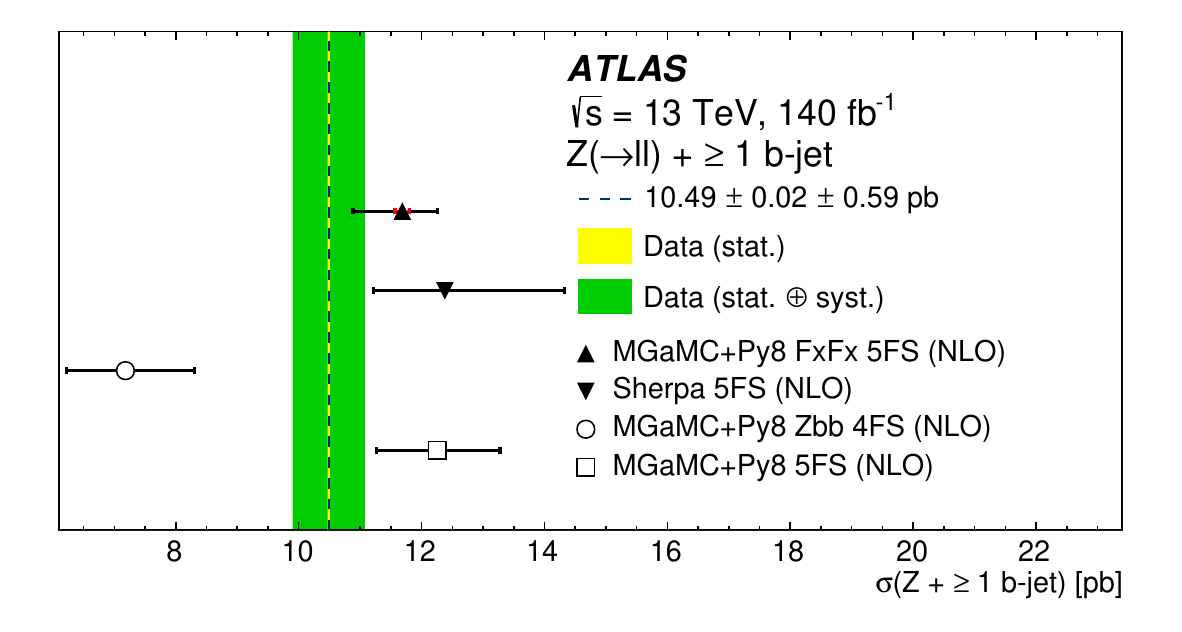}
\caption{}
\end{subfigure}
\begin{subfigure}{0.49\textwidth}
\includegraphics[width=\textwidth]{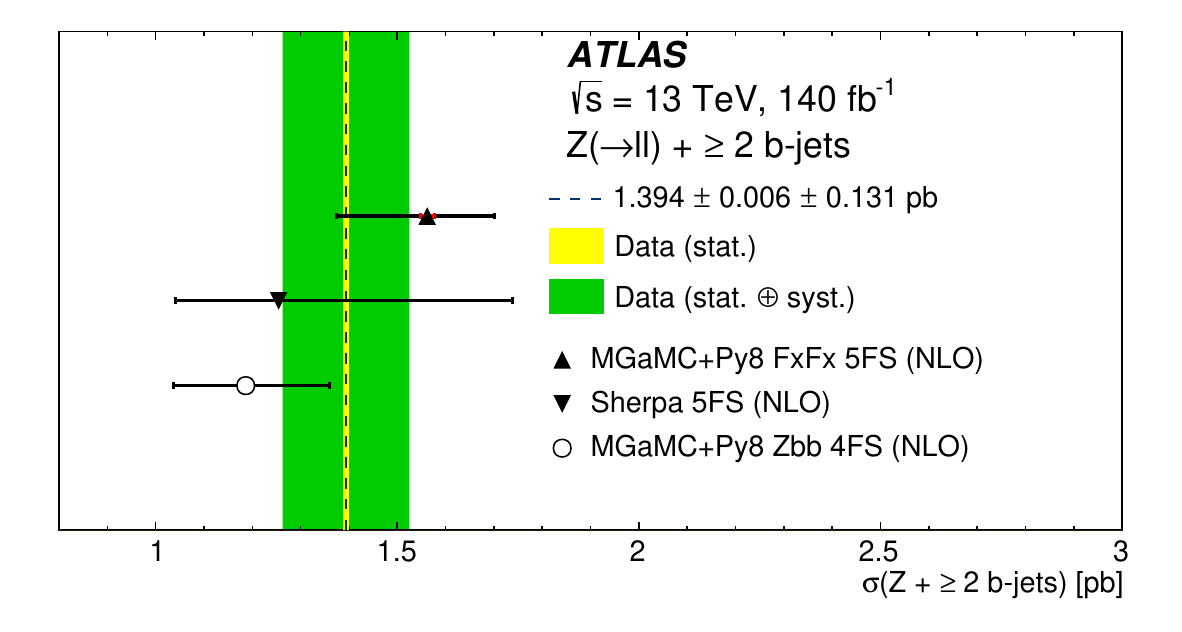}
\caption{}
\end{subfigure}
\end{center}
\caption{
Measured fiducial cross-section for (a) $Z\,+ \ge 1~b$-jet and (b) $Z\,+ \ge 2~b$-jets production. The data are compared with the predictions from
the 5FS multi-leg generators \textsc{MGaMC+Py8 FxFx}  and \textsc{Sherpa 2.2.11}, with \textsc{MGaMC+Py8 4FS (NLO)}, and with \textsc{MGaMC+Py8 5FS (NLO)}.
The thin inner band corresponds to the statistical uncertainty of the data, and the outer band to statistical and systematic uncertainties of the data, added in quadrature.
The error bars on the \textsc{MGaMC+Py8 FxFx}, \textsc{Sherpa 2.2.11}, and \textsc{MGaMC+Py8} predictions correspond to the statistical and theoretical uncertainties added in quadrature. The sum in quadrature of statistical and PDF related uncertainties are shown as inner bars.
}
\label{fig:pred_XS_incl_B}
\end{figure}

\begin{figure}[!tb]
\begin{center}
\includegraphics[width=0.6\textwidth]{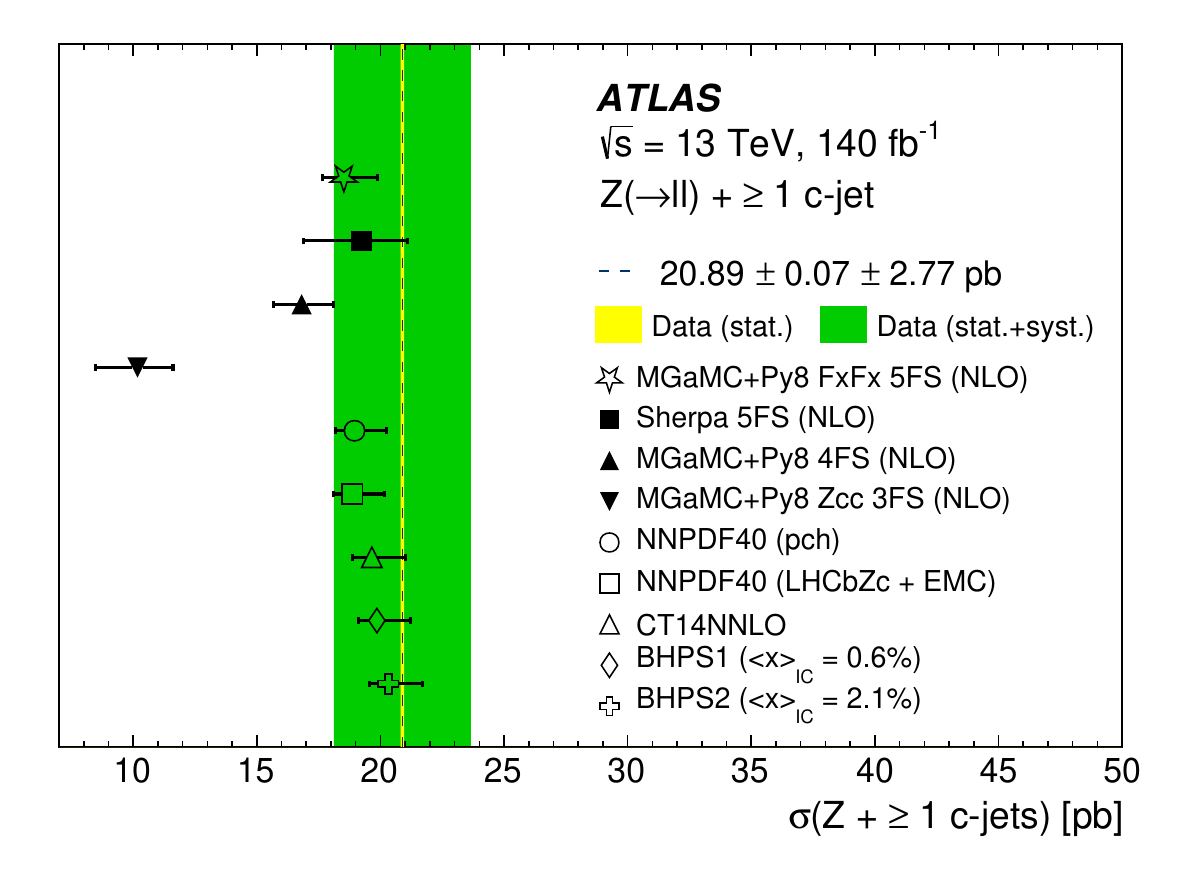}
\end{center}
\caption{
Measured fiducial cross-section for $Z\,+ \ge 1~c$-jet production. The data are compared with the 5FS predictions from
\textsc{MGaMC+Py8 FxFx} and \textsc{Sherpa 2.2.11}, with \textsc{MGaMC+Py8 3FS (NLO)}, and with \textsc{MGaMC+Py8 4FS (NLO)},
and with various PDFs based on different intrinsic charm models (see Section~\ref{sec:datasets}).
The thin inner band corresponds to the statistical uncertainty of the data, and the outer band to statistical and systematic uncertainties of the data, added in quadrature.
The error bars on the \textsc{MGaMC+Py8 FxFx}, and \textsc{Sherpa 2.2.11} predictions correspond to the statistical and theoretical uncertainties added in quadrature.
}
\label{fig:pred_XS_incl_C}
\end{figure}

\FloatBarrier

\subsection{Differential cross-sections for \Zoneb}
\label{sec:results:Z1b}

The differential cross-section measurements for the \Zoneb process are shown in
Figures~\ref{fig:pred_fiducial1B_ZPt_and_bPt}--\ref{fig:pred_fiducial1B_Zb_DR}.
They are compared with the predictions from the 5FS multi-leg generators \mgfxfx and \SHERPA, with \mgamcfour and \mgamcfive, and
with NLO and NNLO fixed-order predictions~\cite{Gauld:2023zlv}.

As the fixed-order predictions are made at parton level and,
moreover, use a different jet flavour definition~\cite{Gauld:2022lem},
the unfolded cross-sections cannot be compared with them directly.
To allow for a proper comparison, two corrections are applied to the predictions.
The first one accounts for the hadronisation and multi-parton-interaction (MPI) effects.
It is calculated as the ratio of the simulated distributions at hadron level
with MPI enabled to those at parton level with MPI disabled,
using a dedicated sample of \Zjets events generated with \PYTHIA[8.310] at LO accuracy.
The jet flavour definition in this case uses the flavour-dressing algorithm~\cite{Gauld:2022lem}.
The second correction accounts for the effect of different jet flavour classification algorithms.
It is calculated as the ratio of hadron-level distributions using the jet flavour definition described
in Section~\ref{sec:unfolding} to those made with the flavour-dressing algorithm, using \mgfxfx sample.
This correction is also derived using the \SHERPA sample and the difference with \mgfxfx is treated as its uncertainty.
Both corrections are applied to the fixed-order predictions as bin-by-bin multiplicative factors.
Uncertainties in these predictions shown on the plot correspond to the sum of the intrinsic theory uncertainty
of the predictions and the total uncertainty of the corrections, while the former are also shown separately.

The distributions of the transverse momentum of the \Zbos and of the \bjets probe pQCD over a wide range of scales  and
provide important input to the background prediction for other SM processes and searches beyond the SM.

The differential cross-section as a function of \ptZ\  for events with at least one \bjet is shown
in Figure~\ref{fig:pred_fiducial1B_ZPt_and_bPt}(a).
In general, the measured spectrum is harder than all predicted spectra. Overall, the prediction from \mgfxfx
demonstrates the best agreement with data, while the \SHERPA prediction still describes the data within larger theory uncertainties.
\mgamcfive predicts a softer \ptZ\ spectrum resulting in an overestimate of  the low-\pT range, while \mgamcfour
is below data in the entire range. The NLO fixed-order computation predicts a noticeably softer spectrum than in data.
This discrepancy reduces with NNLO prediction, however, it still cannot describe the entire spectrum.

Figure~\ref{fig:pred_fiducial1B_ZPt_and_bPt}(b) shows the leading \bjet \pT distribution.
Both \mgfxfx and \SHERPA predictions agree with data within their theory uncertainties.
The \mgamcfive prediction overestimates the data cross-section in the low \pT regime, while for higher \pT values the data are modelled well.
Both NLO and NNLO fixed-order calculations describe well almost the entire spectrum, only slightly
underpredict the data at \pT values above 500~\GeV.

It is noticeable that the uncertainty of the NNLO predictions is dominated by the that of the correction
described above, while their own uncertainty evaluated by varying the QCD scales is small. This indicates
the importance of using IRC-safe jet flavour definitions in future precision measurements of similar processes.

The distribution of \DeltaRZb is sensitive to the presence of additional radiation in the event.
It is shown in Figure~\ref{fig:pred_fiducial1B_Zb_DR}.
Both  \mgfxfx and \SHERPA predictions describe data rather well, only overestimating data
in the  $\DeltaRZb \approx \pi$ region corresponding to the $Z+b$ back-to-back topology.
The same trend is present in the \mgamcfive prediction, with the discrepancy being even larger.
The worst agreement with data is shown by \mgamcfour, where topologies in the tail of the distribution~-- i.e. collinear and at large \DeltaRZb~-- undershoot the data by more than 50\%.
Fixed-order calculations suffer from divergences at $\DeltaRZb \approx \pi$ and their uncertainties are high in that
region. The NLO calculation tends to underestimate the cross-section for \DeltaRZb between 1 and 2.5 and above 4,
but the description becomes almost perfect when moving to NNLO.

\begin{figure}[!tb]
\begin{center}
\begin{subfigure}{0.49\textwidth}
\includegraphics[width=\textwidth]{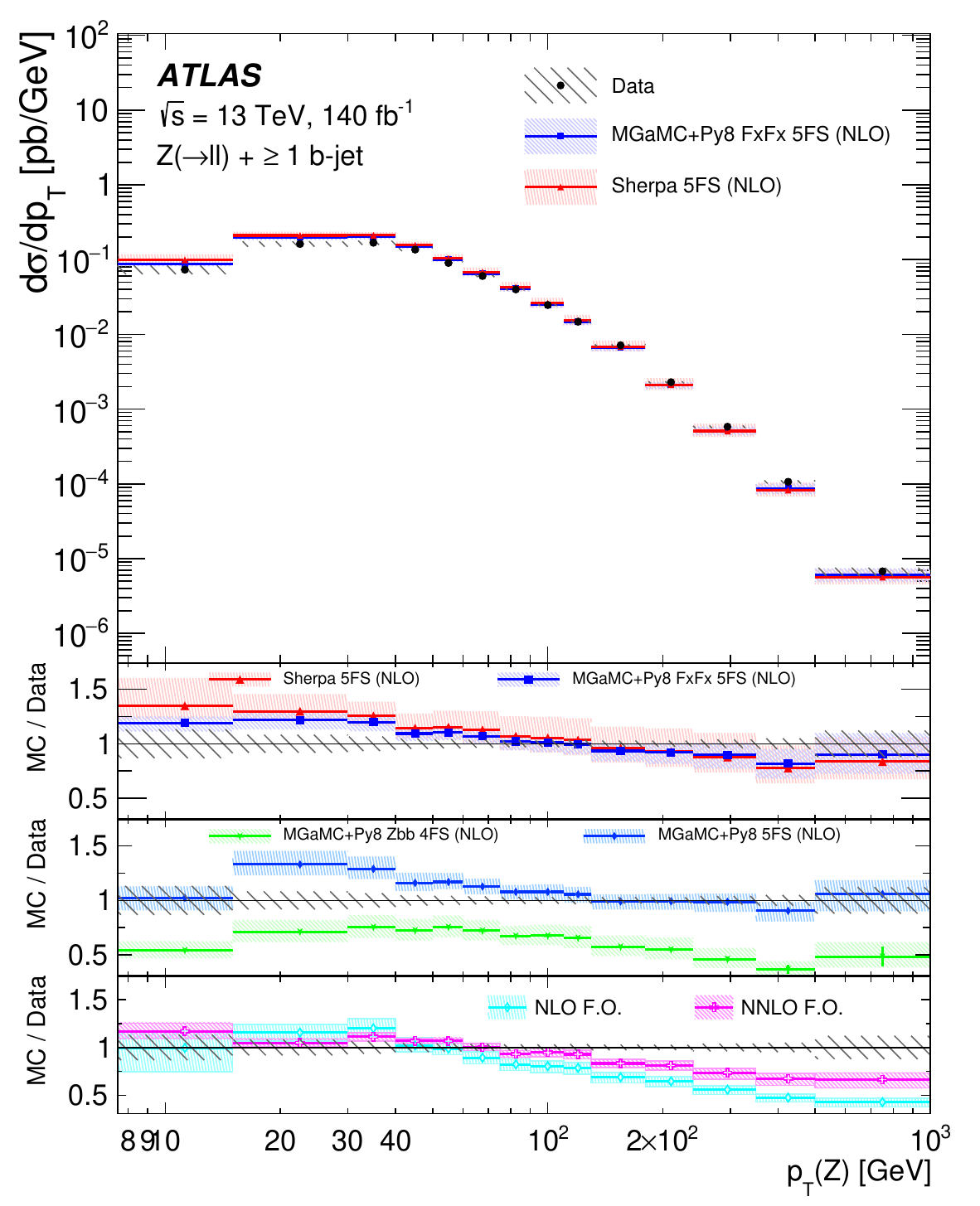}
\caption{}
\end{subfigure}
\begin{subfigure}{0.49\textwidth}
\includegraphics[width=\textwidth]{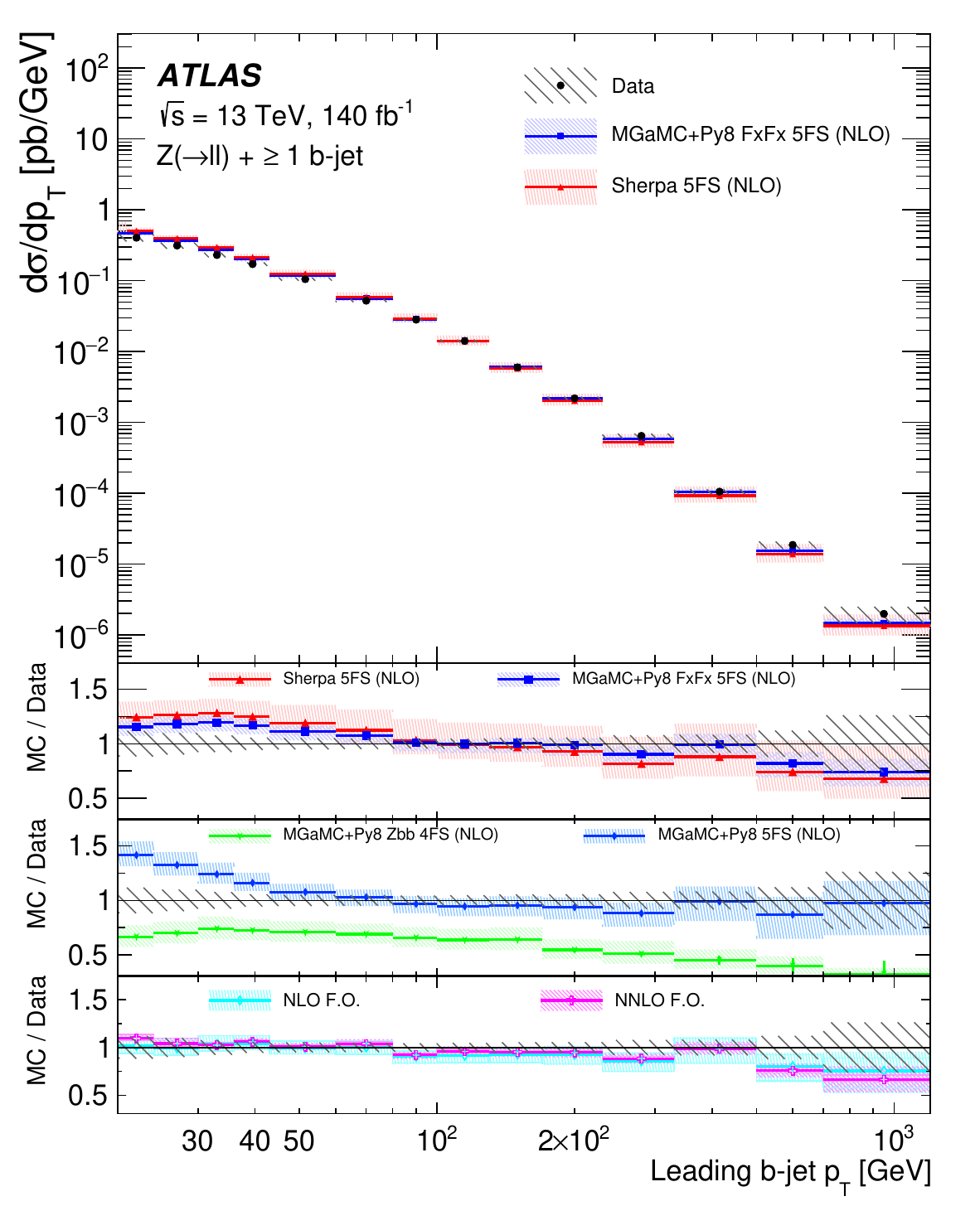}
\caption{}
\end{subfigure}
\end{center}
\caption{
Measured fiducial cross-section for $Z\,+ \ge 1~b$-jet  production as a function of (a) $p_{\textrm{T}}(Z)$  and (b) leading $b$-jet $p_{\textrm{T}}$.
The data are compared with the predictions from
the 5FS multi-leg generators \textsc{MGaMC+Py8 FxFx} and \textsc{Sherpa 2.2.11}, with \textsc{MGaMC+Py8 4FS (NLO)} and \textsc{MGaMC+Py8 5FS (NLO)}, and
with NLO and NNLO fixed-order (F.O.) predictions~\cite{Gauld:2023zlv}.
The error bars correspond to the statistical uncertainty, and the
hatched bands to the data statistical and systematic uncertainties added in quadrature. The shaded bands correspond to
the statistical and theoretical uncertainties of the predictions added in quadrature.
For the fixed-order predictions, the uncertainties in the hadronisation and MPI and flavour definition algorithm corrections
are also added in quadrature to the total, while pure theory uncertainty of the predictions are shown as the range
between the horizontal lines.
}
\label{fig:pred_fiducial1B_ZPt_and_bPt}
\end{figure}

\begin{figure}[!tb]
\begin{center}
\includegraphics[width=.49\linewidth]{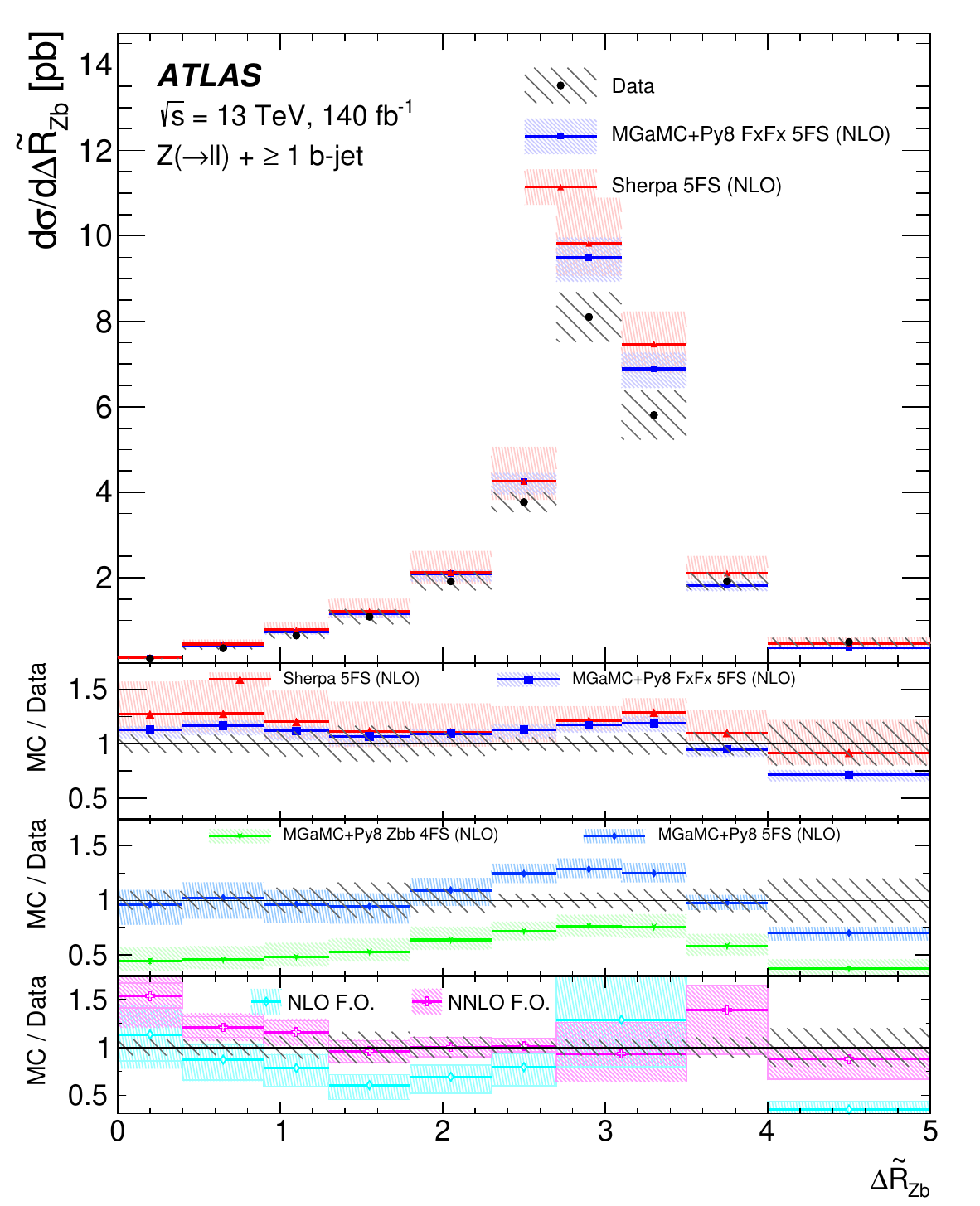}
\end{center}
\caption{
Measured fiducial cross-section for $Z\,+ \ge 1~b$-jet production as a function of $\Delta \tilde R_{Zb}$.
The data are compared with the predictions from
the 5FS multi-leg generators \textsc{MGaMC+Py8 FxFx} and \textsc{Sherpa 2.2.11}, with \textsc{MGaMC+Py8 4FS (NLO)} and \textsc{MGaMC+Py8 5FS (NLO)}, and
with NLO and NNLO fixed-order (F.O.) predictions~\cite{Gauld:2023zlv}.
The error bars correspond to the statistical uncertainty, and the
hatched bands to the data statistical and systematic uncertainties added in quadrature. The shaded bands correspond to
the statistical and theoretical uncertainties of the predictions added in quadrature.
For the fixed-order predictions, the uncertainties in the hadronisation and MPI and flavour definition algorithm corrections
are also added in quadrature to the total, while pure theory uncertainty of the predictions are shown as the range
between the horizontal lines.
}
\label{fig:pred_fiducial1B_Zb_DR}
\end{figure}

\FloatBarrier

\subsection{Differential cross-sections for \Ztwob}

Events with a \Zbos produced in association with two \bjets\ constitute an important background for other SM and beyond-SM processes, such as Higgs boson production in association with a \Zbos or potential new physics signatures with similar final states.
Furthermore, they probe the mechanism of  gluon splitting into \bquarks.
The differential cross-section measurements for \Ztwob are shown in Figure~\ref{fig:pred_fiducial2B}.
They are compared with predictions from
the 5FS multi-leg generators \mgfxfx and \SHERPA and with \mgamcfour.
No fixed-order calculations are provided for \Ztwob observables within the framework used in Ref.~\cite{Gauld:2023zlv}.

A measurement of the  angular separation between the two leading \bjets allows characterisation of the hard radiation
at large angles and the soft radiation for collinear emissions.
The distribution of the azimuthal angle between the two leading \bjets, \DeltaPhibb,
is chosen as an example of such an observable and the corresponding differential cross-section is shown in Figure~\ref{fig:pred_fiducial2B}(a).
All predictions generally agree with data within their theory uncertainties. Both \mgfxfx and \SHERPA describe the
data shape well, while \mgamcfour tends to underestimate small and large \DeltaPhibb values corresponding to
collinear and back-to-back \bjets.

The invariant mass of the two leading \bjets is an important observable in the measurement of
associated $ZH$ production with the Higgs boson decaying into $b\bar{b}$,
and in searches for physics beyond the SM in the same final state.
The differential cross-section as a function of \mbb is shown in Figure~\ref{fig:pred_fiducial2B}(b).
In general, all calculations predict steeper growth below 80~\GeV and steeper decrease for higher values. \mgamcfour provides a good description of
the data in the widest range between 60 and 600~\GeV, but still fails to describe them for smaller and larger \mbb.
\mgfxfx overestimates the data near the maximum of the distribution at 60--100~\GeV and underestimates them above 600~\GeV.
\SHERPA is in agreement with the data within its large uncertainty up to 400~\GeV and deviates from the data  at higher values.

\begin{figure}[!tb]
\begin{center}
\begin{subfigure}{0.49\textwidth}
\includegraphics[width=\textwidth]{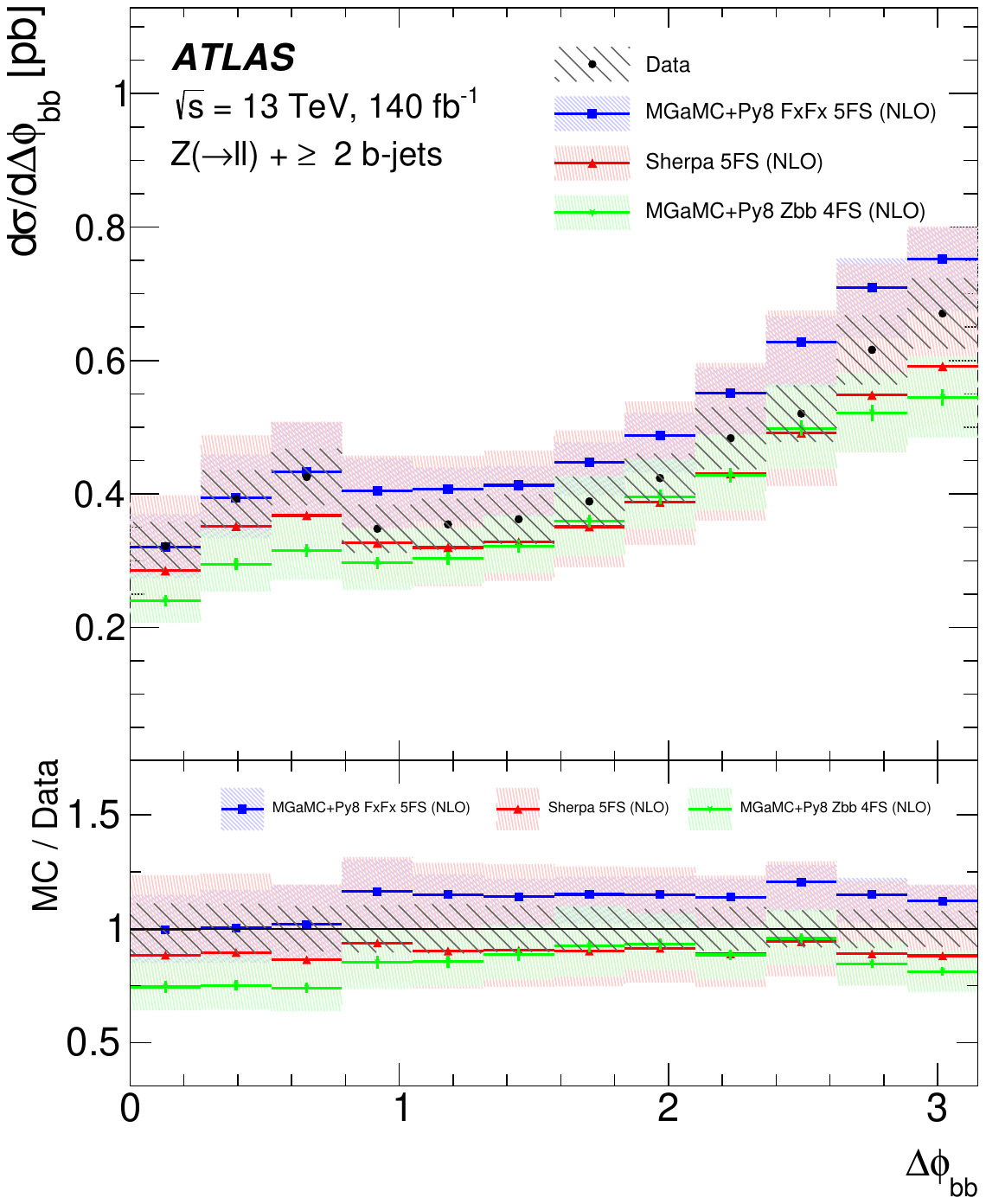}
\caption{}
\end{subfigure}
\begin{subfigure}{0.49\textwidth}
\includegraphics[width=\textwidth]{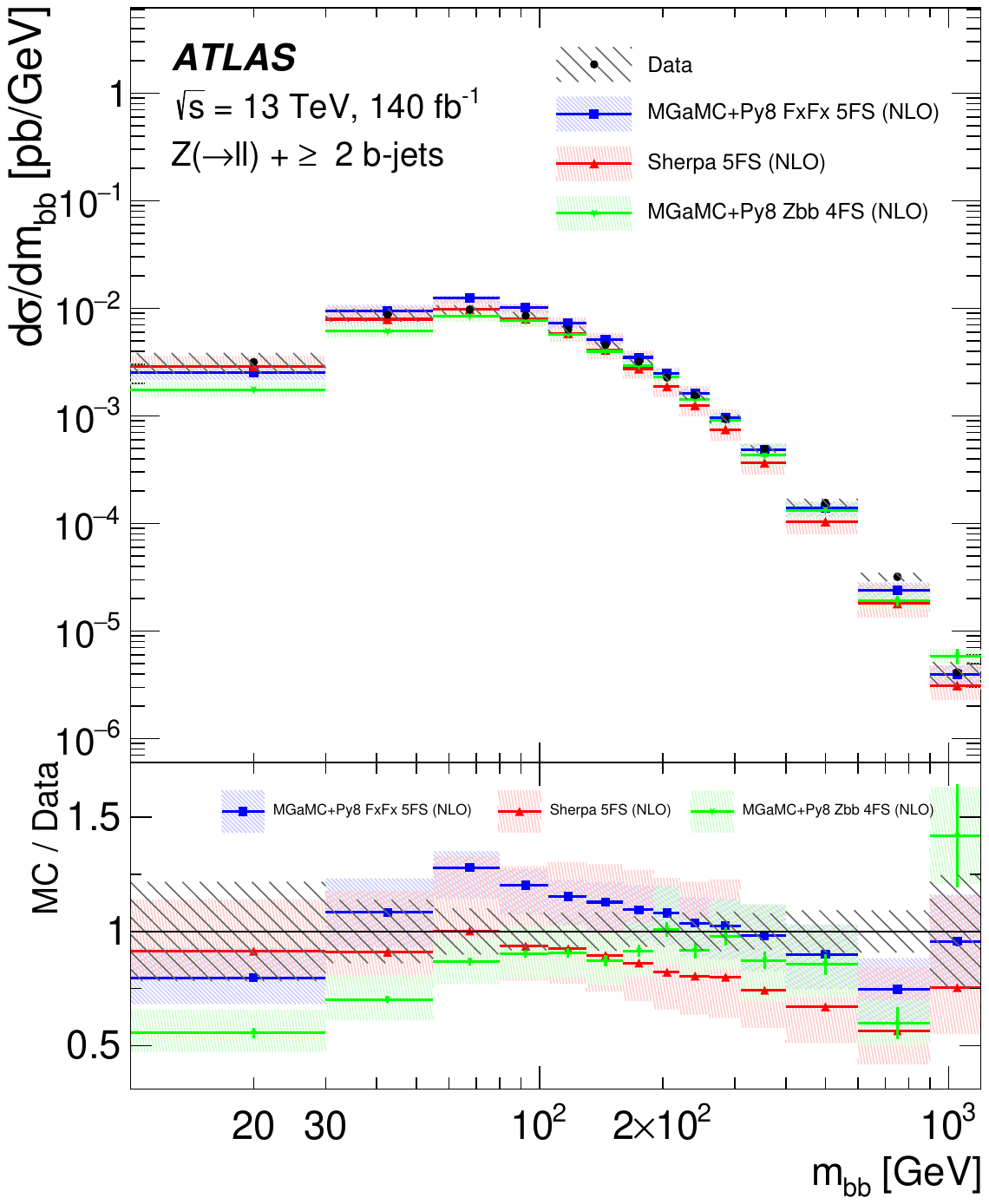}
\caption{}
\end{subfigure}

\end{center}
\caption{
Measured fiducial cross-section for $Z\,+ \ge 2~b$-jets  production as a function of (a) $\Delta\phi_{bb}$ and (b) $m_{bb}$.
The data are compared with the predictions from
the 5FS multi-leg generators \textsc{MGaMC+Py8 FxFx} and \textsc{Sherpa 2.2.11}, and with \textsc{MGaMC+Py8 4FS (NLO)}.
The error bars correspond to the statistical uncertainty, and the
hatched bands to the data statistical and systematic uncertainties added in quadrature. The shaded bands correspond to
the statistical and theoretical uncertainties of the predictions added in quadrature.
}
\label{fig:pred_fiducial2B}
\end{figure}

\FloatBarrier

\subsection{Differential cross-sections for \Zonec}

Differential cross-section measurements for the \Zonec process constitute an important probe of pQCD and of the charm PDF\@.
The results are presented in Figures~\ref{fig:pred_fiducial1C_FS_1}--\ref{fig:pred_fiducial1C_xF_and_Ratio}.
The differential \Zcjets\ cross-section  measurements are compared with the predictions from
the 5FS multi-leg generators \SHERPA and \mgfxfx, with \mgamcthree and \mgamcfour, and
with NLO and NNLO fixed-order predictions~\cite{Gauld:2023zlv}.
The latter are corrected for the effects related to the hadronisation and MPI and to the different jet flavour classification algorithms
as described in Section~\ref{sec:results:Z1b}.
These comparisons are shown in Figures~\ref{fig:pred_fiducial1C_FS_1}--\ref{fig:pred_fiducial1C_FS_2}.

Besides, the measurements are compared to \mgfxfx\ predictions with various PDFs, probing the IC models as
listed in Section~\ref{sec:datasets:theory}. Comparisons to those predictions are shown in
Figure~\ref{fig:pred_fiducial1C_xF_and_Ratio}.

Differential cross-sections as function of \pT of the \Zbos and the leading \cjet are shown in Figure~\ref{fig:pred_fiducial1C_FS_1}.
Both \pT spectra are described well by \mgfxfx and \SHERPA in the soft part, while above 40--50~\GeV (80--100~\GeV) for \Zbos (\cjet) \pT the data cross-section
is significantly underestimated by these predictions. A better description of the data shape overall is provided by \mgamcfour,
however, it is generally near or beyond the lower edge of the data uncertainty band. The \mgamcthree prediction is
significantly below
the data as mentioned above when discussing the inclusive cross-sections.
This discrepancy is noticeably larger than that between the \mgamcfour prediction and the \Zoneb measurement, which  can be attributed to the different
masses of $b$- and \cquarks, causing those logarithmic terms to be larger for the latter.

The NLO fixed-order calculation predicts softer spectra of both \Zbos and leading \cjet \pT than that in data.
The discrepancy is the most noticeable for \ptZ above 50--100~\GeV. Moving to NNLO precision improves the agreement
only slightly for \ptZ and has no impact on \ptclead.

Figure~\ref{fig:pred_fiducial1C_FS_2} shows the distribution of \xF of the leading \cjet. \mgfxfx and \SHERPA predict
a steeper slope of the \xF spectrum compared to the data.
At the same time, the \mgamcfour prediction and both NLO and NNLO fixed-order
calculations describe the data shape well, while systematically underestimating the overall normalization.

\begin{figure}[!tb]
\begin{center}
\begin{subfigure}{0.49\textwidth}
\includegraphics[width=\textwidth]{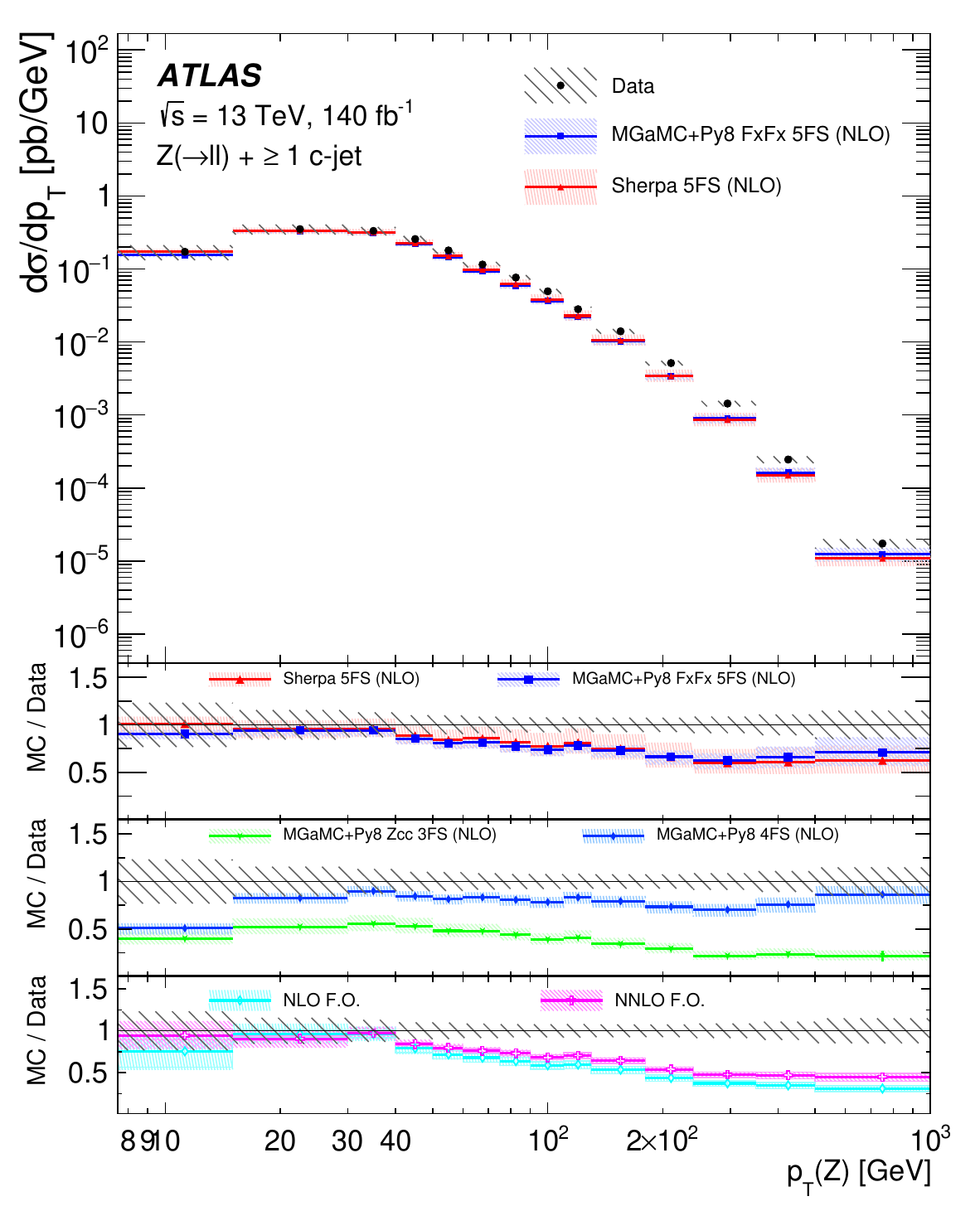}
\caption{}
\end{subfigure}
\begin{subfigure}{0.49\textwidth}
\includegraphics[width=\textwidth]{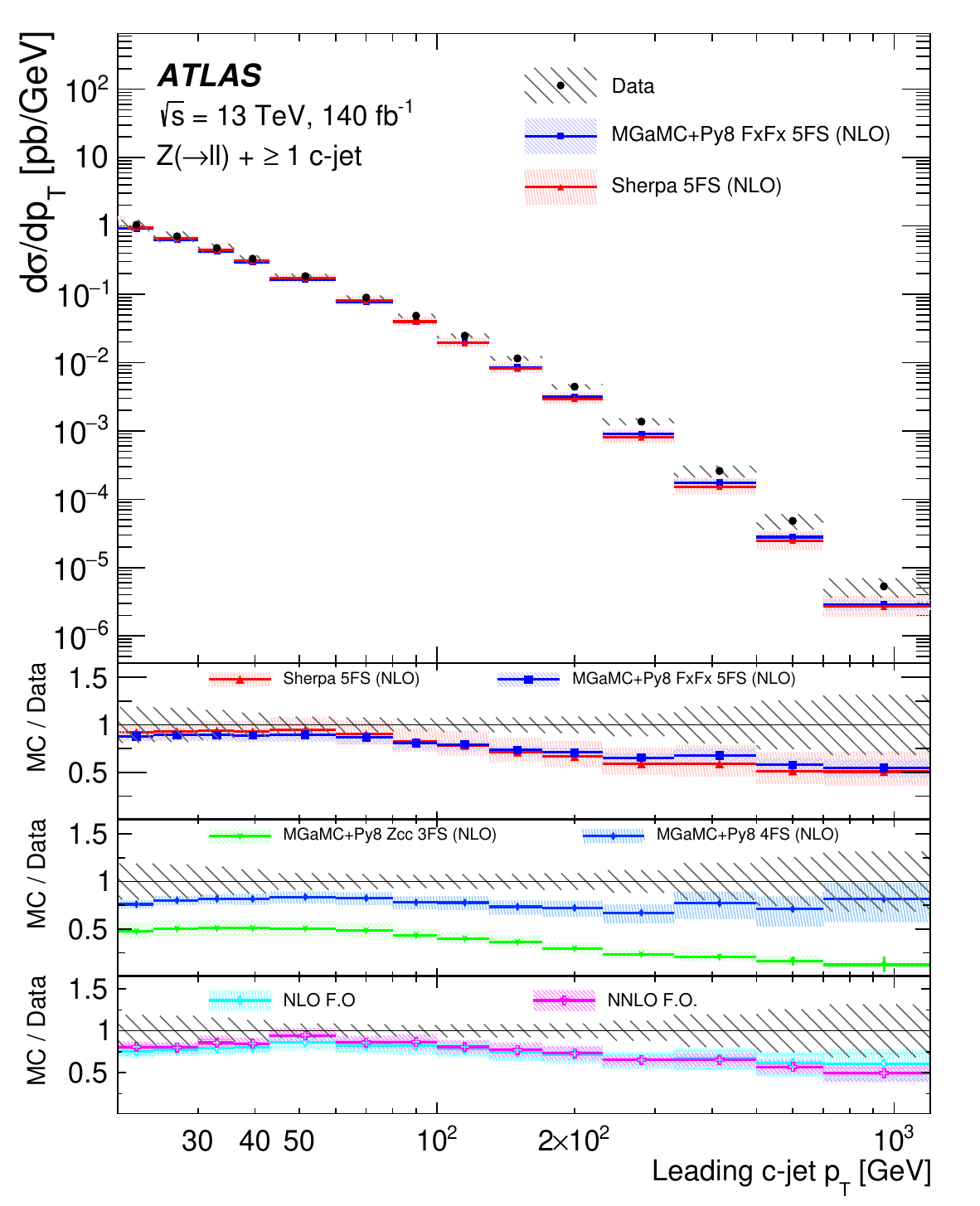}
\caption{}
\end{subfigure}

\end{center}
\caption{
Measured fiducial cross-section for $Z\,+ \ge 1~c$-jet  production as a function of (a) $p_{\textrm{T}}(Z)$  and (b) leading $c$-jet $p_{\textrm{T}}$.
The data are compared with the predictions from
the 5FS multi-leg generators \textsc{MGaMC+Py8 FxFx} and \textsc{Sherpa 2.2.11}, with \textsc{MGaMC+Py8 3FS (NLO)} and \textsc{MGaMC+Py8 4FS (NLO)}, and
with NLO and NNLO fixed-order (F.O.) predictions~\cite{Gauld:2023zlv}.
The error bars correspond to the statistical uncertainty, and the
hatched bands to the data statistical and systematic uncertainties added in quadrature. The shaded bands correspond to
the statistical and theoretical uncertainties of the predictions added in quadrature.
For the fixed-order predictions, the uncertainties in the hadronisation and MPI and flavour definition algorithm corrections
are also added in quadrature to the total, while pure theory uncertainty of the predictions are shown as the range
between the horizontal lines.
}
\label{fig:pred_fiducial1C_FS_1}
\end{figure}

\begin{figure}[!tb]
\begin{center}
\includegraphics[width=.49\linewidth]{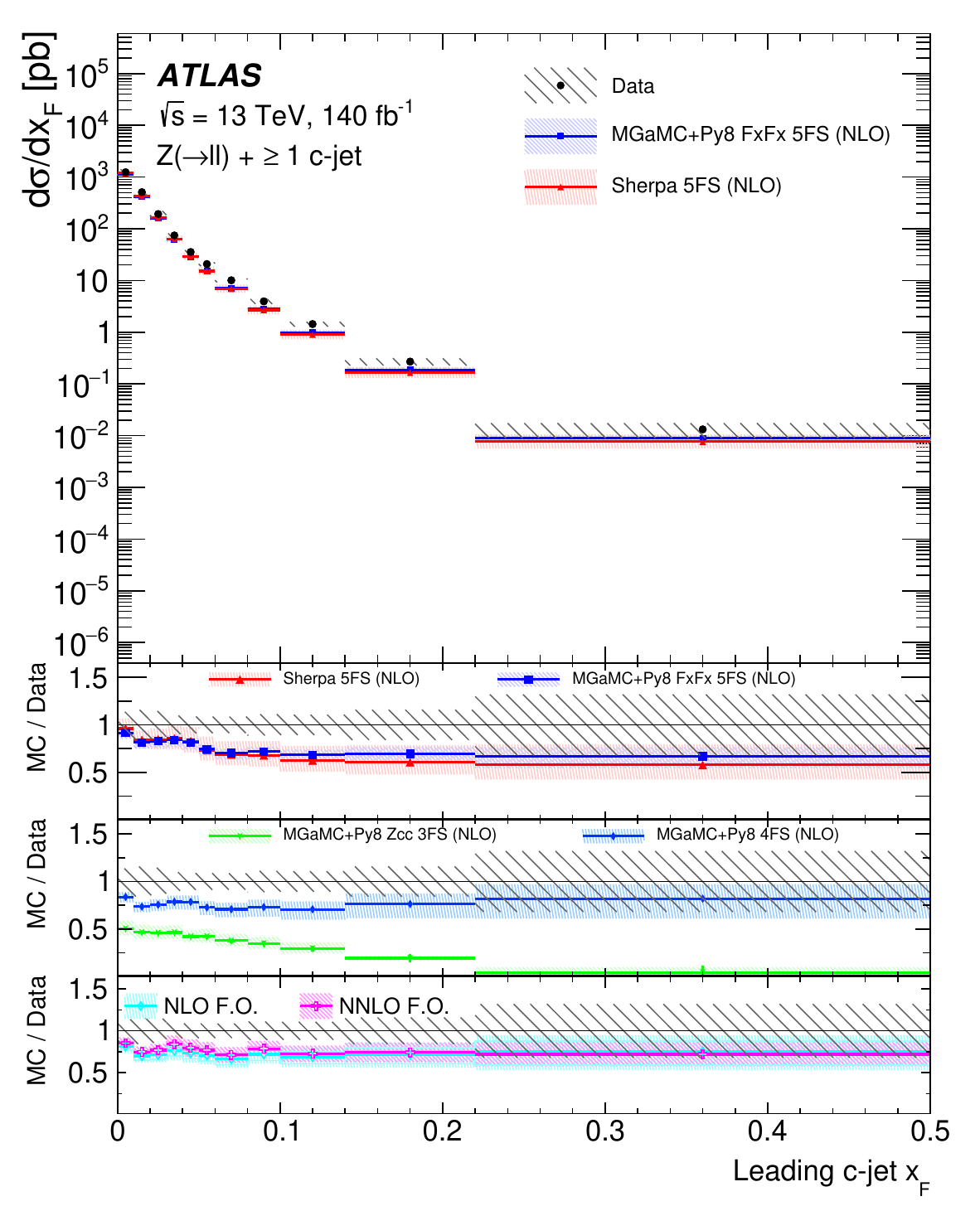}
\end{center}
\caption{
Measured fiducial  cross-section for $Z\,+ \ge 1~c$-jet  production as a function of leading $c$-jet $x_F$.
The data are compared with the predictions from
the 5FS multi-leg generators \textsc{MGaMC+Py8 FxFx} and \textsc{Sherpa 2.2.11}, with \textsc{MGaMC+Py8 3FS (NLO)} and \textsc{MGaMC+Py8 4FS (NLO)}, and
with NLO and NNLO fixed-order (F.O.) predictions~\cite{Gauld:2023zlv}.
The error bars correspond to the statistical uncertainty, and the
hatched bands to the data statistical and systematic uncertainties added in quadrature. The shaded bands correspond to
the statistical and theoretical uncertainties of the predictions added in quadrature.
For the fixed-order predictions, the uncertainties in the hadronisation and MPI and flavour definition algorithm corrections
are also added in quadrature to the total, while pure theory uncertainty of the predictions are shown as the range
between the horizontal lines.
}
\label{fig:pred_fiducial1C_FS_2}
\end{figure}

Figure~\ref{fig:pred_fiducial1C_xF_and_Ratio} presents comparisons of the measured cross-section as a function of
leading \cjet \xF and of  \RptZ for events with at least one \cjet with various IC models.
The \xF distribution is more sensitive to the IC contribution.
The large amount of IC in the BHPS2 model noticeably improves the agreement to data.
However, for more realistic IC scenarios, such as BHPS1 and the IC PDFs from the NNPDF and CT18 families, the improvement is still marginal, being
noticeable only in the last 1--2 bins.
In the measurement of \RptZ the effect of IC is more significant compared to both experimental and theory uncertainties, thanks to their significant cancellation in this ratio.
However, only the BHPS2 model has a non-negligible effect on the agreement to data, making it slightly better in
some bins but worse in others compared to the predictions with no-IC CT14NNLO PDF\@.
Other IC PDFs from NNPDF and CT18 families do not make significant difference in modelling these observables.

\begin{figure}[!tb]
\begin{center}
\begin{subfigure}{0.49\textwidth}
\includegraphics[width=\textwidth]{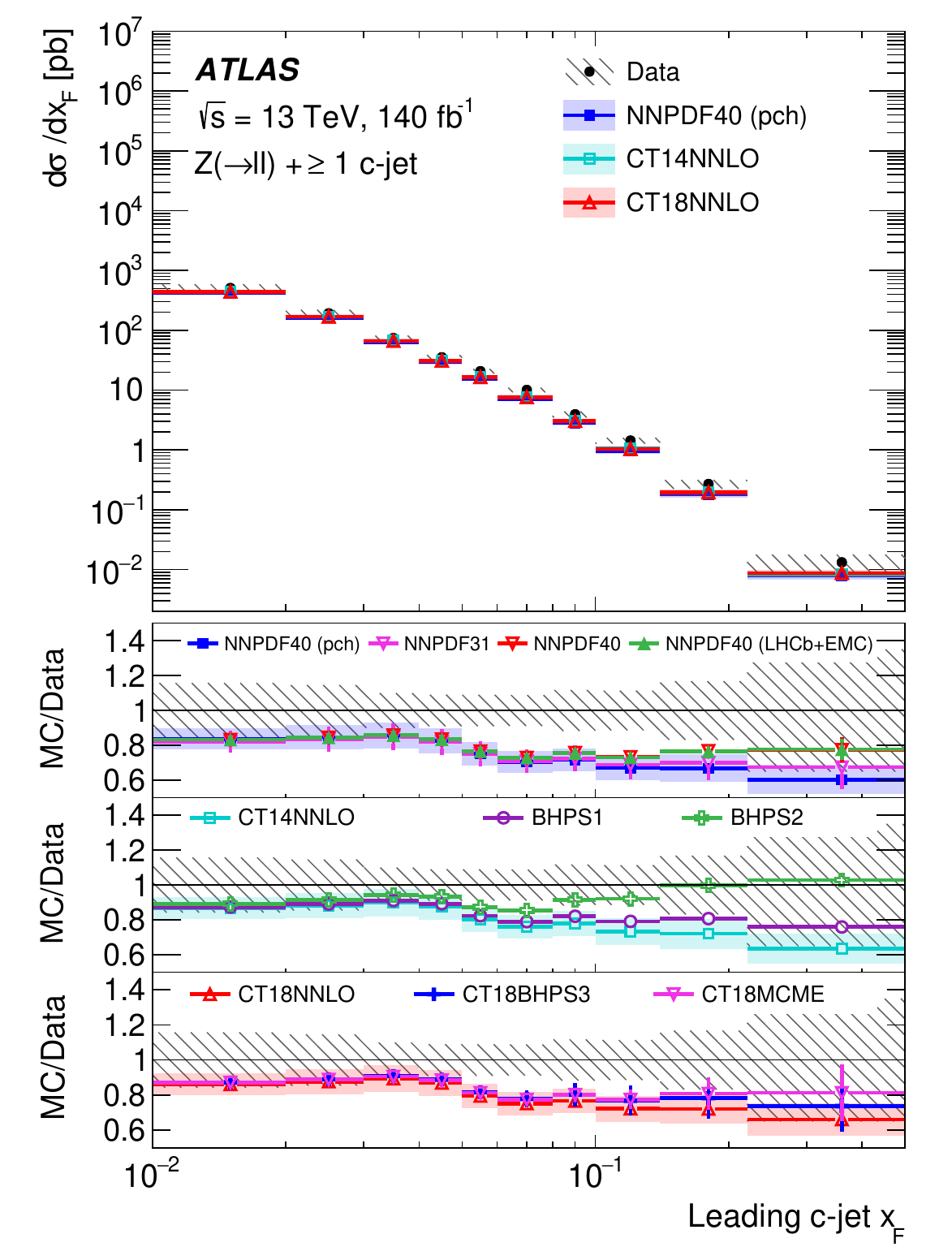}
\caption{}
\end{subfigure}
\begin{subfigure}{0.49\textwidth}
\includegraphics[width=\textwidth]{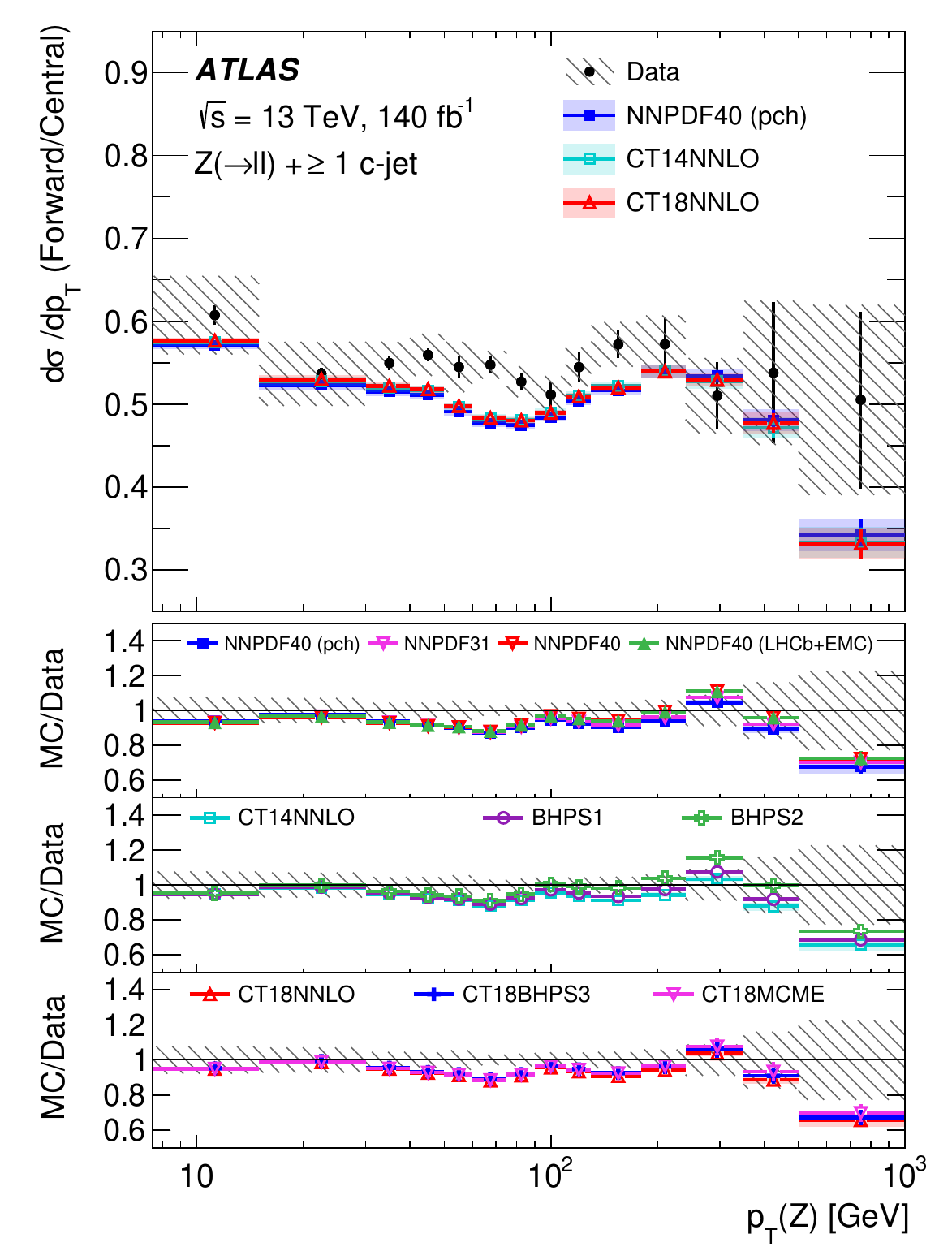}
\caption{}
\end{subfigure}
\end{center}
\caption{
Measured fiducial cross-section for $Z\,+ \ge 1~c$-jet  production as a function of (a)  leading $c$-jet $x_F$  and  (b) R($p_{\textrm{T}}(Z)$).
The data are compared with the nominal \textsc{MGaMC+Py8 FxFx}  predictions and with those using the PDFs testing several IC models.
The error bars correspond to the statistical uncertainty, and the
hatched bands to the data statistical and systematic uncertainties added in quadrature. The shaded bands correspond to
the statistical and theoretical uncertainties of the predictions added in quadrature.
}
\label{fig:pred_fiducial1C_xF_and_Ratio}
\end{figure}


\clearpage

\FloatBarrier

\section{Conclusion}
\label{sec:conclusion}

This paper presents a measurement of the production rate of a \Zbos in association with jets originating from \bquarks\ and \cquarks\
(\Zbjets and \Zcjets) in proton--proton collisions at $\sqrt{s}=13$~\TeV, using data corresponding to an integrated luminosity of 140~\ifb collected by the ATLAS experiment at the CERN LHC.

The cross-sections are measured using the electron and muon decay modes of the \Zbos in a fiducial phase space.
Inclusive cross-sections in a fiducial phase space are measured for events with at least one or at least two \bjets\ and for events with at least one \cjet.
Differential cross-sections are measured as a function of \pT of the leading \bjet,
the \pT of the \Zbos and the difference in angular separation between the \Zbos\ and the leading \bjet\
for events with at least one \bjet. Measurements are also performed as a function of the \pT of the leading \cjet,
the \pT of the \Zbos, and the Feynman-$x$ variable \xF for events with at least one \cjet.
A separate measurement is performed of the  ratio of
\Zcjets production cross-sections as a function of the \pT of the \Zbos measured in two ranges of \Zbos rapidity, central and forward.
Finally, the differential cross-sections are measured as a function of the invariant mass of the two leading \bjets\ and as a function
of the azimuthal angle difference between the two leading \bjets\ for events with at least two \bjets.

Measurements of the \Zcjets\ production cross-section are performed for the first time at the ATLAS experiment
while the \Zbjets production measurements significantly improve on the precision of the previous ATLAS results~\cite{STDM-2017-38}.
Both types of measurements significantly extend to wider kinematic ranges compared to similar
measurements performed at $\sqrt{s}=13$~\TeV by other experiments~\cite{CMS-SMP-19-011,CMS-SMP-20-015}.

The measurements are compared with predictions from a variety of Monte Carlo generators.
In general, the 5FS \mgfxfx\ and \SHERPA\ predictions describe the \Zbjets\ data within the experimental and theory uncertainties whereas the 5FS \SHERPA\ predictions show larger theory uncertainties than \mgfxfx.
The \mgamcZbbNLO\ generator systematically underestimates the \Zoneb\ distributions and does not provide a good description of the angular separation between the \Zbos\ and the leading \bjet,
but describes the \Ztwob\ cross-sections.
All generators underestimate the \Zcjet\ cross-sections: the 5FS \mgfxfx\ and \SHERPA\ predictions underestimate the data for medium ranges of \Zbos\ and \cjet\ \pT, whereas the \mgamcZccNLO\ predictions  systematically underestimate all kinematic \Zcjet\ regions by a large factor.
The 5FS NNLO \Zoneb\ predictions with flavour dressing show a similar performance as the 5FS multi-leg \mgfxfx\ and SHERPA\ generators but predict a
softer \bjet\ \pT spectrum than the data. The 5FS NNLO \Zcjet\ predictions underestimate the data cross-sections even more that their 5FS multi-leg counterparts.
\mgfxfx\ versions with PDFs with different IC content were compared to \Zcjet\ distributions and for a given observable no significant difference in the modelling between the various PDFs was found.

This measurement provides important input for the improvement of theoretical predictions and Monte Carlo generators of \Zbos production in association with \bjets and \cjets, allowing a better quantitative understanding of perturbative QCD.


\clearpage

\section*{Acknowledgements}

%
%

%
%

We thank CERN for the very successful operation of the LHC and its injectors, as well as the support staff at
CERN and at our institutions worldwide without whom ATLAS could not be operated efficiently.

The crucial computing support from all WLCG partners is acknowledged gratefully, in particular from CERN, the ATLAS Tier-1 facilities at TRIUMF/SFU (Canada), NDGF (Denmark, Norway, Sweden), CC-IN2P3 (France), KIT/GridKA (Germany), INFN-CNAF (Italy), NL-T1 (Netherlands), PIC (Spain), RAL (UK) and BNL (USA), the Tier-2 facilities worldwide and large non-WLCG resource providers. Major contributors of computing resources are listed in Ref.~\cite{ATL-SOFT-PUB-2023-001}.

We gratefully acknowledge the support of ANPCyT, Argentina; YerPhI, Armenia; ARC, Australia; BMWFW and FWF, Austria; ANAS, Azerbaijan; CNPq and FAPESP, Brazil; NSERC, NRC and CFI, Canada; CERN; ANID, Chile; CAS, MOST and NSFC, China; Minciencias, Colombia; MEYS CR, Czech Republic; DNRF and DNSRC, Denmark; IN2P3-CNRS and CEA-DRF/IRFU, France; SRNSFG, Georgia; BMBF, HGF and MPG, Germany; GSRI, Greece; RGC and Hong Kong SAR, China; ISF and Benoziyo Center, Israel; INFN, Italy; MEXT and JSPS, Japan; CNRST, Morocco; NWO, Netherlands; RCN, Norway; MNiSW, Poland; FCT, Portugal; MNE/IFA, Romania; MESTD, Serbia; MSSR, Slovakia; ARRS and MIZ\v{S}, Slovenia; DSI/NRF, South Africa; MICINN, Spain; SRC and Wallenberg Foundation, Sweden; SERI, SNSF and Cantons of Bern and Geneva, Switzerland; MOST, Taipei; TENMAK, T\"urkiye; STFC, United Kingdom; DOE and NSF, United States of America.

Individual groups and members have received support from BCKDF, CANARIE, CRC and DRAC, Canada; PRIMUS 21/SCI/017, CERN-CZ and FORTE, Czech Republic; COST, ERC, ERDF, Horizon 2020, ICSC-NextGenerationEU and Marie Sk{\l}odowska-Curie Actions, European Union; Investissements d'Avenir Labex, Investissements d'Avenir Idex and ANR, France; DFG and AvH Foundation, Germany; Herakleitos, Thales and Aristeia programmes co-financed by EU-ESF and the Greek NSRF, Greece; BSF-NSF and MINERVA, Israel; Norwegian Financial Mechanism 2014-2021, Norway; NCN and NAWA, Poland; La Caixa Banking Foundation, CERCA Programme Generalitat de Catalunya and PROMETEO and GenT Programmes Generalitat Valenciana, Spain; G\"{o}ran Gustafssons Stiftelse, Sweden; The Royal Society and Leverhulme Trust, United Kingdom.

In addition, individual members wish to acknowledge support from CERN: European Organization for Nuclear Research (CERN PJAS); Chile: Agencia Nacional de Investigaci\'on y Desarrollo (FONDECYT 1190886, FONDECYT 1210400, FONDECYT 1230812, FONDECYT 1230987); China: Chinese Ministry of Science and Technology (MOST-2023YFA1605700), National Natural Science Foundation of China (NSFC - 12175119, NSFC 12275265, NSFC-12075060); Czech Republic: Czech Science Foundation (GACR - 24-11373S), Ministry of Education Youth and Sports (FORTE CZ.02.01.01/00/22\_008/0004632), PRIMUS Research Programme (PRIMUS/21/SCI/017); EU: H2020 European Research Council (ERC - 101002463); European Union: European Research Council (ERC - 948254, ERC 101089007), Horizon 2020 Framework Programme (MUCCA - CHIST-ERA-19-XAI-00), European Union, Future Artificial Intelligence Research (FAIR-NextGenerationEU PE00000013), Italian Center for High Performance Computing, Big Data and Quantum Computing (ICSC, NextGenerationEU); France: Agence Nationale de la Recherche (ANR-20-CE31-0013, ANR-21-CE31-0013, ANR-21-CE31-0022), Investissements d'Avenir Labex (ANR-11-LABX-0012); Germany: Baden-Württemberg Stiftung (BW Stiftung-Postdoc Eliteprogramme), Deutsche Forschungsgemeinschaft (DFG - 469666862, DFG - CR 312/5-2); Italy: Istituto Nazionale di Fisica Nucleare (ICSC, NextGenerationEU); Japan: Japan Society for the Promotion of Science (JSPS KAKENHI JP21H05085, JSPS KAKENHI JP22H01227, JSPS KAKENHI JP22H04944, JSPS KAKENHI JP22KK0227); Netherlands: Netherlands Organisation for Scientific Research (NWO Veni 2020 - VI.Veni.202.179); Norway: Research Council of Norway (RCN-314472); Poland: Polish National Agency for Academic Exchange (PPN/PPO/2020/1/00002/U/00001), Polish National Science Centre (NCN 2021/42/E/ST2/00350, NCN OPUS nr 2022/47/B/ST2/03059, NCN UMO-2019/34/E/ST2/00393, UMO-2020/37/B/ST2/01043, UMO-2021/40/C/ST2/00187, UMO-2022/47/O/ST2/00148, UMO-2023/49/B/ST2/04085); Slovenia: Slovenian Research Agency (ARIS grant J1-3010); Spain: Generalitat Valenciana (Artemisa, FEDER, IDIFEDER/2018/048), Ministry of Science and Innovation (MCIN \& NextGenEU PCI2022-135018-2, MICIN \& FEDER PID2021-125273NB, RYC2019-028510-I, RYC2020-030254-I, RYC2021-031273-I, RYC2022-038164-I), PROMETEO and GenT Programmes Generalitat Valenciana (CIDEGENT/2019/023, CIDEGENT/2019/027); Sweden: Swedish Research Council (Swedish Research Council 2023-04654, VR 2018-00482, VR 2022-03845, VR 2022-04683, VR 2023-03403, VR grant 2021-03651), Knut and Alice Wallenberg Foundation (KAW 2018.0157, KAW 2018.0458, KAW 2019.0447, KAW 2022.0358); Switzerland: Swiss National Science Foundation (SNSF - PCEFP2\_194658); United Kingdom: Leverhulme Trust (Leverhulme Trust RPG-2020-004), Royal Society (NIF-R1-231091); United States of America: U.S. Department of Energy (ECA DE-AC02-76SF00515), Neubauer Family Foundation.
We thank R.\,Gauld (MPI Munich) and the NNLOJET developers for providing NLO $Z+b$ 5FS and $Z+c$ 4FS predictions with \mgamc as well as NLO and NNLO fixed-order predictions~\cite{Gauld:2023zlv} in the fiducial phase space of the analysis, and for helpful discussions.

%
%


%
%
%
%
%
%
%
%

%
%
%
%
%
%
%
%

%
%
%
\printbibliography
\clearpage
 
\begin{flushleft}
\hypersetup{urlcolor=black}
{\Large The ATLAS Collaboration}

\bigskip

\AtlasOrcid[0000-0002-6665-4934]{G.~Aad}$^\textrm{\scriptsize 103}$,
\AtlasOrcid[0000-0001-7616-1554]{E.~Aakvaag}$^\textrm{\scriptsize 16}$,
\AtlasOrcid[0000-0002-5888-2734]{B.~Abbott}$^\textrm{\scriptsize 121}$,
\AtlasOrcid[0000-0002-0287-5869]{S.~Abdelhameed}$^\textrm{\scriptsize 117a}$,
\AtlasOrcid[0000-0002-1002-1652]{K.~Abeling}$^\textrm{\scriptsize 55}$,
\AtlasOrcid[0000-0001-5763-2760]{N.J.~Abicht}$^\textrm{\scriptsize 49}$,
\AtlasOrcid[0000-0002-8496-9294]{S.H.~Abidi}$^\textrm{\scriptsize 29}$,
\AtlasOrcid[0009-0003-6578-220X]{M.~Aboelela}$^\textrm{\scriptsize 44}$,
\AtlasOrcid[0000-0002-9987-2292]{A.~Aboulhorma}$^\textrm{\scriptsize 35e}$,
\AtlasOrcid[0000-0001-5329-6640]{H.~Abramowicz}$^\textrm{\scriptsize 152}$
% ...
\bigskip
\\

% ...
$^{16}$Department for Physics and Technology, University of Bergen, Bergen; Norway.\\
% ...
$^{29}$Physics Department, Brookhaven National Laboratory, Upton NY; United States of America.\\
% ...
$^{35}$$^{(a)}$Facult\'e des Sciences Ain Chock, R\'eseau Universitaire de Physique des Hautes Energies - Universit\'e Hassan II, Casablanca;$^{(b)}$Facult\'{e} des Sciences, Universit\'{e} Ibn-Tofail, K\'{e}nitra;$^{(c)}$Facult\'e des Sciences Semlalia, Universit\'e Cadi Ayyad, LPHEA-Marrakech;$^{(d)}$LPMR, Facult\'e des Sciences, Universit\'e Mohamed Premier, Oujda;$^{(e)}$Facult\'e des sciences, Universit\'e Mohammed V, Rabat;$^{(f)}$Institute of Applied Physics, Mohammed VI Polytechnic University, Ben Guerir; Morocco.\\
% ...
$^{44}$Physics Department, Southern Methodist University, Dallas TX; United States of America.\\
% ...
$^{49}$Fakult\"{a}t Physik , Technische Universit{\"a}t Dortmund, Dortmund; Germany.\\
% ...
$^{55}$II. Physikalisches Institut, Georg-August-Universit\"{a}t G\"ottingen, G\"ottingen; Germany.\\
% ...
$^{103}$CPPM, Aix-Marseille Universit\'e, CNRS/IN2P3, Marseille; France.\\
% ...
$^{117}$$^{(a)}$New York University Abu Dhabi, Abu Dhabi;$^{(b)}$United Arab Emirates University, Al Ain; United Arab Emirates.\\
% ...
$^{121}$Homer L. Dodge Department of Physics and Astronomy, University of Oklahoma, Norman OK; United States of America.\\
% ...
$^{152}$Raymond and Beverly Sackler School of Physics and Astronomy, Tel Aviv University, Tel Aviv; Israel.\\
% ...
$^{a}$ Also Affiliated with an institute covered by a cooperation agreement with CERN.\\
$^{b}$ Also at An-Najah National University, Nablus; Palestine.\\
$^{c}$ Also at Borough of Manhattan Community College, City University of New York, New York NY; United States of America.\\
$^{d}$ Also at Center for Interdisciplinary Research and Innovation (CIRI-AUTH), Thessaloniki; Greece.\\
$^{e}$ Also at Centro Studi e Ricerche Enrico Fermi; Italy.\\
$^{f}$ Also at CERN, Geneva; Switzerland.\\
% ...

\end{flushleft}


%

%
%

%

%
%
%
%

\end{document}